# Cooperative ligation breaks sequence symmetry and stabilizes early molecular replication

Shoichi Toyabe[1, 2] and Dieter Braun[1]

[1] Systems Biophysics, Physics Department, NanoSystems Initiative Munich and Center for Nanoscience, Ludwig-Maximilians-Universität München, Amalienstrasse 54, 80799 München, Germany; [2] Department of Applied Physics, Graduate School of Engineering, Tohoku University, Aramaki-aza Aoba 6-6-05, Aoba-ku, Sendai 980-8579, Japan

**Abstract.**

*Each living species carries a complex DNA sequence that determines their unique features and functionalities. It is generally assumed that life started from a random pool of oligonucleotides sequences, generated by a prebiotic polymerization of nucleotides. The mechanism that initially facilitated the emergence of sequences that code for the function of the first species from such a random pool of sequences remains unknown. It is a central problem of the origin of life. An interesting option would be a self-selection mechanism by spontaneous symmetry breaking. Initial concentration fluctuations of specific sequence motifs would have been amplified and outcompeted less abundant sequences, enhancing the signal to noise to replicate and select functional sequences. Here, we demonstrate with experimental and theoretical findings that templated ligation would provide such a self-selection. In templated ligation, two adjacent single sequences strands are chemically joined when a third complementary strand sequence brought them in close proximity. This simple mechanism was a likely side-product of a prebiotic polymerization chemistry once the strands reach the length to form double stranded species. As shown here, the ligation gave rise to a nonlinear replication process by the cooperative ligation of matching sequences which self-promoted their own elongation. This led to a cascade of enhanced template binding and faster ligation reactions. A requirement was the reshuffling of the strands by thermal cycling, enabled for example by microscale convection. By using a limited initial sequence space and performing long term ligations, we found that complementary sequences with an initially higher concentration prevailed over either non-complementary or less concentrated sequences. The latter died out by the molecular degradation that we simulated in the experiment by serial dilution. The experimental results are consistent with both explicit and abstract theory models that were generated considering the ligation rates determined experimentally. Previously, other nonlinear modes of replication such as Hypercycles have been discussed to overcome instabilities from first-order replication dynamics such as the error catastrophe and the dominance of structurally simple, but fast replicating sequences, known as the Spiegelman problem. Assuming that templated ligation was driven by the same chemical mechanism that generated prebiotic polymerization of oligonucleotides, the mechanism could function as a missing link between polymerization and the self-stabilized replication, offering a pathway to the autonomous emergence of Darwinian evolution for the origin of life.*

The genetic information of present-day living species is encoded in the DNA sequence as a specific combination of four different nucleotides A, C, G, and T. How the first functional information coding sequences could have emerged from an initial pool of random sequences is one of the central questions to understand the origin of life.

Any sequence space of even moderate length of for example 25 bases is so large ($4^{25} \approx 10^{15}$) that even with a significant volume and concentration, the sampling can only be sparse, meaning that each molecule would have a different sequence. It must be expected that even if such a very short sequence would have encoded and conferred an advantageous function for molecular evolution, it would not have had an impact on such a random pool of sequences. This limited sampling of sequence space is even the case for the most sophisticated Systematic Evolution of Ligands by Exponential Enrichment (SELEX) lab procedures that are intended to select for functional molecules despite the fact that human brains, hands and complex machines are guiding the evolutionary process.

Spontaneous symmetry breaking is a basic physical mechanism that creates structure out of random initial conditions. We assessed here whether nonlinear effects in a basic replication mechanism could implement the symmetry breaking in sequence space. Stochastic fluctuations in the initial sequence distribution would be amplified and give rise to an increasing homogeneous sequence space at a given location (Figure 1). With the same process happening at different locations with stochastic variations, a large diversity of sequences could be sampled. Each sequence would be present at each location in significant concentrations such that a selection based on its function could be implemented. To test this scenario, we studied how under the replication dynamics of templated ligation, sets of similar sequences with high concentrations could survive by replication while less concentrated or uncorrelated sequences die out.

It is no coincidence that a replication mechanism is the means to drive the self-selection dynamics of spontaneous symmetry breaking. Replication is necessary to later maintain and improve the functional sequence by the mechanism of Darwinian evolution. Therefore, the mechanism could be a route towards the emergence of sequence species for Darwinian evolution.

The understanding of the emergence of life has advanced and progressed fast in the last years. Examples include big steps forward in RNA catalyzed replication[1-8], synthesis of nucleotides[9,10] and base-by-base replication with activated nucleotides[11,12]. Autocatalytic replication of sequence information by the catalytic function of ribozymes is thought to be crucial for the evolution of central functions of biology. Autocatalytic replication has been demonstrated with carefully designed and selected ribozymes[13,14], where exponential growth of a group of mutually-catalytic ribozymes was observed. However, the search of mechanisms allowing the spontaneous emergence of such complex autocatalytic sequences from a pool of random sequences remains an unsolved problem.

The first information molecules did not have many mechanisms at their disposal. After the synthesis and accumulation of the first nucleotides, random sequences[15-19] could polymerize. Once they were long enough to bind at the given temperature, three-molecule complexes form, and one sequence would bind to two complementary sequences which can then be connected by a suitable chemical reaction. This three-body reaction is termed a templated ligation. It offers a most basic replication mechanism since the two sequences are linked only if the sequences match sufficiently, offering a transfer of information from one to another molecule (Figure 1).

Our experiments revealed a reduction of the sequence space by replicating sequences with templated ligation. It is based on the fact that longer sequences are more likely to bind, and this replicates faster with the templated ligation. In addition, matching sequences can increase their length by ligation and therefore enhance their replication speed. As discussed with the experiments and theoretical descriptions below, this finding makes the replication nonlinear with respect to sequence concentrations.

In addition to the intractable search through a huge sequence space, early replicators faced two major instabilities: the error catastrophe[20] and a convergence towards the fastest replicators, also known as the Spiegelman problem[21]. The first sequences are shaped by the balance between mutation and selection in a fitness landscape[20,22,23]. If the error rate exceeds a certain threshold, the selection can no longer suppress the accumulation of sequence errors and the sequences vanishes. This is termed an error catastrophe. This would have been a major bottleneck for early molecular evolution since primitive replicators would initially have had only a limited fidelity. The dilemma is that strands require long and structured sequences in order to provide the necessary catalytic activity to decrease the error rate. But, for longer sequences, it was much harder to keep the error rate below the error threshold of the error catastrophe. Therefore, even if a self-replicating molecule did emerge at some point in a soup of random sequences, it would have been difficult for it to replicate only its own sequence information against the sequence majority of the pool.

The second dilemma is that first-order exponential growth tends to converge to a sequence with the highest replication rate. However, the faster replication of ever shorter mutant sequences was very likely. Therefore, the inherent shortening of the strands by mutations likely suppressed the possibility to use oligonucleotides for the storage of sufficient information unless another process would have enhanced the length of the sequences. In many experimental systems, the shortest possible sequence will end up dominating the population, as experimentally shown by Spiegelman[21,24], and also recently observed for the case of Ribo-PCR by Joyce[7]. Both, the error catastrophe and the convergence to a common sequence would have strongly limited the emergence of early molecular Darwinian evolution.

The hypercycle proposed by Eigen and Schuster is a theoretical concept that can overcome both dilemmas[20]. A hypercycle is a ring-shaped network of replication reactions in which the product of a replication cycle catalyzes the reaction of the following replication cycle. This cooperative mechanism amplifies the sequence information at a second order[18]; $\dot{x} = kx^2 - dx$. Here, $k$ and $d$ are the replication and the deletion rates which can differ between replicators. Unlike the first-order growth, the growth rate of the hypercycle $\dot{x}/x = kx - d$ depends on $x$ thus, it is enhanced with the accumulation of $x$. Sexual reproduction is another example of a higher-order growth[25]. That is, two elements react to generate the next generation of the elements. This has been shown in vitro with an enzymatic DNA-RNA amplification system called cooperative amplification of templates by cross-hybridization (CATCH)[26,27]. The frequency-dependent selection, or the Allee effect[20], caused by a nonlinear growth of sequences could stabilize the wild types and could raise the error threshold. With the frequency-dependent selection, even replicators with smaller $k$ could survive and dominate if they once obtained a high frequency by fluctuations.

## Results

### Templated ligation.

The implementation of templated ligation experiments of DNA strands required first, a ligation mechanism that could facilitated the formation of a phosphodiester bond between the 3'-hydroxyl and 5'-phosphate groups of two adjacent DNA strands. And second, the presence of a third template DNA strand that by complementary base-pairing could bring the strands in close proximity for the ligation (Figure S1).

Prebiotically plausible molecules that can carry out such ligation reaction, in an efficient and rapid way, have not yet been identified. In this regard, diamidophosphate (DAP) was reported[28] as a potential prebiotic candidate molecule that could favored the oligomerization/condensation of nucleotides and amino acids into their respective polymers. However, the slow kinetic and low yield of such processes would need the support of physical non-equilibrium boundary conditions.

Faster methods and strategies exist, for example, the use of 1-ethyl-3-(3-dimethylaminopropyl) carbodiimide (EDC) as *in-situ* activator of phosphate groups of nucleotides but the reaction suffers from modifications of the oligonucleotides at elevated temperatures. Furthermore, EDC in its optimized form is prebiotically not very plausible, but it can trigger both the polymerization and templated ligation of RNA or DNA.

Therefore, we decided to use a highly evolved protein to ligate the DNA strands. The ligation reactions experiments were conducted with a thermostable Taq DNA ligase[29] that catalyzes the ligation reaction >100-fold faster than EDC. We used elevated temperatures and relatively long sequences to reduce the impact of a sequence dependence in the catalytic activity of the ligase protein. In order to mathematically model the system, the sequence space was reduced to three different 20mer DNA sequences denoted with lower case letters, **a** = 5'-atcag gtgga agtgc tggtt, **b** = 5'-atgag ggaca aggca acagt and **c** = 5'-attgg gtcac atcgg agtct and their reverse complements $\overline{a}$, $\overline{b}$ and $\overline{c}$ with a single-base overhang at the 5' end to avoid blunt-end ligations. Capital letters denote both the respective sequences and their complements A = {**a**, $\overline{a}$}, B = {**b**, $\overline{b}$} and C = {**c**, $\overline{c}$}. The sequences were designed to have comparable ligation rates, similar melting temperatures and reduced self-annealing (supplement S1). To allow the shuffling between hybridized strands prior to ligation, all experiments were conducted under the following thermal cycling conditions (67 °C for 10 s and 95 °C for 5 s). Such thermal cycling could have been provided by thermal microscale convection[24,30] in prebiotic environments.

### Length dependence of templated ligation.

Under the abovementioned conditions, 20mer sequences were found to bind less stably than 40mers or 60mer oligomers to the same complementary 60mer template sequence and therefore longer sequences were elongated faster than shorter sequences (Figure 2a,b,c). Template ligations were performed under thermal cycling with the template sequence $\overline{cba}$. The kinetics of the ligation were measured for the following three sequence substrate sets: (a) **a** + **b**, (b) **a** + **bc**, and the competitive situation (c) **a** + **b** + **bc** starting with the concentrations [a] = 100 nM, [b] = [bc] = 33.3 nM, and [cba] = 0.25 nM. For the last case (c), the shorter strands **a** + **b** competed with the longer strands **a** + **bc** for ligation on the template. Under the competitive conditions (c), **a** + **bc** ligated to form **abc** with a rate of 215 pM per cycle, 40-fold higher than the 5 pM per cycle rate calculated in order to obtain the elongation product from the shorter strands to form the sequence **ab**. This competitive speed advantage of longer sequences will lead to nonlinear replication.

Before going into the details of the cooperative ligation networks, we will discuss how these ligation experiments were used to infer the experimental parameters of the subsequent models. As we will see, the initial growth rates of ligation products provided both the dissociation constants $K_D$ of strand hybridization and the ligation rates $k$. To measure the initial growth rates $v_{ab}$, $v_{abc}$, $v'_{ab}$, and $v'_{ab}$ for [ab] in (a), [abc] in (b), [ab] and [abc] in (c), respectively, the product concentrations [x](t) were fitted by a simple saturation curve [x](t) = ($v_x$ / q) [1 – exp(- qt)] to find the initial growth rate $v_x$ at t = 0 and the kinetics of saturation q (Figure 2 and supplementary Figure S5). The saturation was caused by the inhibitory binding of the ligation products **ab** in (a) and (c) or **abc** in (b) and (c) to the template $\overline{cba}$. As seen by the measurement results, the found saturation was minimal.

The thermal cycling was sufficiently fast such that the hybridization did not equilibrate during the cycles. Therefore, the experimental data were modeled by assuming that an effective dissociation constant $K_D$ defines the probability of binding under such fast thermal cycling conditions and that the hybridized strands were then ligated with a rate $k$. For the simplest case (a), the growth rate of the product **ab** was thus given by

$$\frac{d[ab]_0}{dt} = k_{ab} \frac{[a]_f [b]_f [\overline{cba}]_f}{K_{D,20} K_{D,20}} \quad (1)$$

Here, [ ]$_f$ denotes the amount of free, unhybridized strands and [ ]$_0$ indicates the total amount of strands. To infer the free strand concentrations, the conservation laws for the binding of three strands are given by:

$$[a]_0 = [a]_{\text{f}} + \frac{[a]_{\text{f}}[\overline{cba}]_{\text{f}}}{K_{\text{D},20}} + \frac{[a]_{\text{f}}[b]_{\text{f}}[\overline{cba}]_{\text{f}}}{K_{\text{D},20}K_{\text{D},20}}, \ [b]_0 = [b]_{\text{f}} + \frac{[b]_{\text{f}}[\overline{cba}]_{\text{f}}}{K_{\text{D},20}} + \frac{[a]_{\text{f}}[b]_{\text{f}}[\overline{cba}]_{\text{f}}}{K_{\text{D},20}K_{\text{D},20}},$$

$$[\overline{cba}]_0 = [\overline{cba}]_{\text{f}} + \frac{[a]_{\text{f}}[b]_{\text{f}}[\overline{cba}]_{\text{f}}}{K_{\text{D},20}K_{\text{D},20}} + \frac{[a]_{\text{f}}[\overline{cba}]_{\text{f}}}{K_{\text{D},20}} + \frac{[b]_{\text{f}}[\overline{cba}]_{\text{f}}}{K_{\text{D},20}} + \frac{[ab]_{\text{f}}[\overline{cba}]_{\text{f}}}{K_{\text{D},40}}. \quad (2)$$

By solving these equations for the three cases (a,b,c) at $t = 0$, we obtained

$$K_{\text{D},20} = \alpha K_{\text{D},40} \text{ and } K_{\text{D},40} = \frac{\beta - 1}{\alpha - \beta} c_0, \quad (3)$$

where $\alpha = v_{\text{abc}} / v_{\text{ab}}$ and $\beta = v'_{\text{abc}} / v'_{\text{ab}}$ were obtained from the experiments.

The ligation rate $k$ was determined with the following reasoning. At small $t$, $[ab]_{\text{f}} = 0$, $[a]_{\text{f}} \approx [a]_0$, and $[b]_{\text{f}} \approx [b]_0$ because the template concentration $[\overline{cba}]_0 = 0.25$ nM was considerably smaller than $[a]_0 = 100$ nM and $[b]_0 = 33$ nM. By inserting these values, we obtained the reaction rates for the ligation of **ab** from **a** + **b** and that of **abc** from **a** + **bc**:

$$k_{\text{ab}} \simeq \frac{v_{\text{ab}}}{[\overline{cba}]_0}\left(1 + \frac{K_{\text{D},20}}{[a]_0} + \frac{K_{\text{D},20}}{[b]_0}\right) \text{ and } k_{\text{abc}} \simeq \frac{v_{\text{abc}}}{[\overline{cba}]_0}\left(1 + \frac{K_{\text{D},20}}{[a]_0} + \frac{K_{\text{D},40}}{[bc]_0}\right) \quad (4)$$

This resulted in the effective dissociation constants for the short and the long strands with $K_{\text{D},20} = 193$ nM, and $K_{\text{D},40} = 4.5$ nM. The ligation rate instead, was found to be $k_{\text{ab}} = 3.0$ nM$^{-1}$cycle$^{-1}$ and $k_{\text{abc}} = 3.0$ nM$^{-1}$cycle$^{-1}$. Since the ligation rate was therefore independent of the oligonucleotide strand length, we assumed $k$ to be constant for all the sequences including substrates and templates. For the case (b) and the competitive case (c) shown in Figure 2, the binding of the ligated product **abc** on the template resulted in the saturation of the growth curves. By fitting the simulated curves with the parameters obtained above, we estimated $K_{\text{D},60}$ to be 2.7 nM. This allowed us to model the ligations in the subsequent simulations with the effective dissociation constants $K_{\text{D},20} = 193$ nM, $K_{\text{D},40} = 4.5$ nM and $K_{\text{D},60} = 2.7$ nM for 20-base, 40-base and 60-base oligomers under the applied temperature cycling.

### *Ligation chain reaction under thermal cycling.*

So far, the elongation sequence product concentration was observed to increase linearly since the complementary short strands were not added to the ligation reactions. The exponential product formation becomes important when for instance the pressure of a serial dilution, simulating the degradation of strands, will exponentially remove molecules by, for example, the diffusion of molecules from the system. We therefore decided to implement a ligation chain reaction with exponential product formation by providing to the ligation reactions **a**, **b**, **c**, and also their complementary sequences $\overline{c}$, $\overline{b}$, $\overline{a}$, as described in Figure 3a. In this case, binding was not expected to be inhibited by the exponential growth due to the periodical temperature cycling and the sufficient supply of the abovementioned shorter strands.

Importantly, if the ligation reaction would not reach an exponential phase, the replicates would have been gradually removed by serial dilution and die out. The molecule degradation present in early molecular evolution was in our experiments approximated by serial dilution. This worst-case approach avoided the possible complexities that could arise from the recycling of degraded sequences.

### *Cooperative ligation network.*

Multiple ligase chain reactions, sharing overlapping sequences, could generate a cooperative ligation network. In the example shown in Figure 3b, the 40-base sequences depicted in the grey circle, AB = {**ab**, $\overline{ba}$}, CA = {**ca**, $\overline{ac}$} and BC = {**bc**, $\overline{cb}$} when supplied with the short 20-base sequence substrates A = {**a**, $\overline{a}$}, B = {**b**, $\overline{b}$} and C = {**c**, $\overline{c}$} can initially experience an exponential but slow growth, due to their short length and low stability to bind the substrates fragments with $K_{\text{D},20}$. However, the sequences AB and BC can cooperate by complementary hybridizing at the common sequence B and elongate giving rise to the novel 60-base (yellow circle). Since the longer ABC sequence can now act as a template for longer overlapping sequences, the ligations AB + C $\rightarrow$ ABC and A + BC $\rightarrow$ ABC show a faster ligation kinetics since $K_{\text{D},40} \ll K_{\text{D},20}$ and provide significantly stronger binding. For this enhanced ligation reaction, the 40-base sequence substrates were provided by feeding from the slower ligation chain reactions with two letters (black arrows). But in addition, the creation of the three letter 60mer sequences also offered an enhanced template to create AB and BC from A, B and C (blue circle). This combination of feed forward to elongate the sequence (yellow) and feed backward to enhance the creation of substrates (blue) increased the concentrations of 40-base and 60-base sequences for cooperating sequences which were able to elongate. As documented with the following experiments, the increase in both complexity and overall ligation kinetics of the growing network therefore increased its replication speed as it progressed outward and gave rise to a concentration dependent, higher order replication dynamics.

*Simple cooperative ligation.*

To test a simple cooperative ligation network, we started the reaction with either a pair of two 40-base sequences that could (AB and BC) or that could not cooperate (AB and AC) in an elongation step with a common sequence pattern. The resulting elongated products were measured with gel electrophoresis (Figure 4a and supplement S4).

For the non-cooperative sequences AB and AC (Figure 4a, left), the logarithmic plot showed an initially exponential growth of the sequences AB and AC. In comparison, the cooperative sequence pair AB and BC (Figure 4a, right) grew similarly in the first 40 cycles, but then the 60-base sequence (ABC) was found to grow at a faster rate. The longer sequence ABC could act as an efficient template for the enhanced ligation rate of producing more AB and BC template (A + BC → ABC; AB + C → ABC). This finding was confirmed by theoretical modeling as indicated below (Figure 4, solid lines).

The concentrations of 40-base, two-letter sequence motifs, denoted by two capital letters in brackets (<AA>,< AB>,< AC>…) were determined by COLD PCR. For example, the concentration <AB> would indicate the concentrations of all motifs AB in all present sequences. In Figure 4, for example, <AB> would correspond to the sum of concentrations of the sequences *ab*, *abc*, *ba*, and *cba*. This was possible experimentally by applying a quantitative PCR method with an initial low denaturation temperature (COLD PCR, further details are provided in supplement S2)[27] which amplified only 40mers, even from longer strands. Note that deep sequencing would not have provided a comparable dynamic concentration range in terms of concentration to record motif concentrations between 0.1 pM and 100 nM needed to measure the sequence dynamics in our experiments. The method was calibrated with known 40mer sequences.

When applied to the experiments in Figure 4a, we found the positive feedback by the elongating sequences. The motifs <AB> and <BC> grew about 2-fold more efficiently for the cooperative starting sequences as composed to the non-cooperative pair <AB> and <AC> (Figure 4b). The cooperative ligation provided a significant boost in the presence of the common sequence motif <AB> and <BC> of the cooperating strands and provided more template concentration to grow more cooperating 40mers from the supplied 20mer sequences.

**Cooperative ligation model.**

The deterministic rate equation systems automatically with a Visual C#-source code. Subsequently, the experiments were modeled on Mathematica 10.1 (Wolfram research, IL, USA) with the obtained deterministic rate equations. The equations used the kinetic rates determined by the experiments in Figure 2. Moreover, the equations took into consideration the binding by hybridization between strands. Conservation laws were used to infer the unbound molecule concentrations and to consider the serial dilutions to feed the substrates and dilute the products. Because of the inherent symmetry in the systems using both (*a*, *b*, *c*) and their complementary sequences ($\bar{a}, \bar{b}, \bar{c}$) in equal amounts, the rate equations were formulated in a simplified manner considering the concentrations of *A*, *B* and *C*. An overview of the modeling method is given below. Further details of the modeling are provided in supplement S5.

As already discussed, thermal cycling is not modeled explicitly, but we considered the experimentally determined ligation rate *k* and effective dissociation constants $K_D$ (Figure 2). In Figure 4 and 5, because the strands did not elongate longer than 60 bases due to the limited number of initial 40-base template strands, we simulated all possible sequences. For the other experiments, it was checked that we could limit the maximum strand length to 120 bases for Figure 5a, to 640 bases for Figure 6b and to 80 bases for Figure 7 without affecting the simulation results significantly. We considered only binding complexes with up to three strands (Figure 6, 7) or four strands (Figure 4 and 5). We estimated that the chance for higher order complexes was much smaller than the three- or four-body complexes.

Despite the reduction of the complexity of the system and the use of a limited sequence space composed of only *a*, *b*, *c* sequences to study the dynamics, the resulting kinetic equation systems reached ASCII-files for Mathematica with a size of around 30 MBytes. We included the Visual C#-source code that was used to generate the equations in the supplement. In order to understand the approach, we show here the logic for the short reaction system of the experiment in Figure 4.

As discussed, when measuring the ligation rates in Figure 2, we determined the effective dissociation constants $K_{D,n}$ experimentally under thermal cycling with an *n*-base overlap. For an overlap longer than 60 bases, we used the same value as $K_{D,60}$. Qualitatively this was a reasonable assumption based on the stronger binding thermodynamics, but a slower binding kinetics due to the smaller concentrations of the binders. We used the following nomenclature to calculate concentrations of binding complexes on the assumption that the hybridization is effectively at equilibrium:

$$[A/AB] = \frac{[A]_f [AB]_f}{K_{D,20}}, \quad [AB, C/ABC] = \frac{[AB]_f [C]_f [ABC]_f}{K_{D,40} K_{D,20}}, \quad [AB, C/A, BC] = \frac{[AB]_f [C]_f [A]_f [BC]_f}{K_{D,20} K_{D,20} K_{D,20}}. \quad (5)$$

Here, [ ]$_f$ denoted the concentration of the free strand. For example, [*AB*, *C* / *A*, *BC*] denoted a four-body complex where AB and BC were hybridized with overhangs, and C and A were hybridized to these overhangs. Some complexes had isomers, for example, AA could hybridize to AAC in two ways, one with only 20-base overlapping and one with a 40-base overlap and no overhang, both requiring different dissociation constants. Partially unbound complexes were included. For example, ABC/ACC had a 40-base overlap in total, therefore, we used $K_{D,40}$ for this hybridization:

$$[ABC/ACC] = \frac{[ABC]_f[ACC]_f}{K_{D,40}}, [AB,C/BBC] = \frac{[AB]_f[C]_f[BBC]_f}{K_{D,20}K_{D,20}}. \tag{6}$$

The following conservation laws were added to the above hybridization expressions to obtain the total concentrations [XY] on the left side from the free concentrations such as $[A]_f$ and $[AB]_f$. The occurrence of a given strand was collected from all the bound complexes. This would lead for example to equations such as

$$[A] = [A]_f + [A/A] + [A/AB] + [A/CA] + [A/BAC] + \cdots + [A,B/AB] + [A,BC/ABC] + [C,A/CBA] + \cdots \tag{7}$$

$$[AB] = [AB]_f + [A/AB] + [AB/AB] + [A/CA] + \cdots + [A,B/AB] + [AB,C/ABC] + 2[AB,AB/ABAB] + \cdots$$

The ligation rate $k$ was not found to be dependent of strand length (Figure 2). Please note that ligations were only modeled if both end sequences of the substrates (B and C in this case) were matched with the template sequences and untemplated ligation was neglected. For example, we obtained:

$$[\dot{A}] = -k'(2[A,A/AA] + [A,B/AB] + [A,BC/ABC] + [B,A/BA] + \cdots)$$

$$[\dot{AB}] = k'([A,B/AB] + [A,B/CAB] + [A,B/ABC] + [A,B/ABB] + \cdots)$$
$$\qquad -k'([AB,C/ABC] + [AB,C/AB] + [A,AB/AA] + \cdots) \tag{8}$$

$$[\dot{ABC}] = k'([AB,C/ABC] + [AB,C/AB] + [AB,C/BBC] + [A,BC/ABC] + [A,BC/AB] + [A,BC/CAB] + \cdots)$$
$$\qquad -k'([ABC,C/CC] - [A,ABC/AAB] + \cdots)$$

The reaction rate $k'$ was only slightly modified with $k'(t) = \alpha k f_n(t)$ from the experimentally determined rate of $k$ = 3.0 nM$^{-1}$cycle$^{-1}$. The coefficient α was necessary to model a saturation due to a limiting amount of the ligase. We used $\alpha$ = 1 in Fig. 4, $\alpha$ = 1.27 in Fig. 5, and $\alpha$ = 1.05 in Fig. 6. We used larger $\alpha$ where the total concentrations of ligating complexes were low compared to that of Fig. 2. The term $f_t(t)$ modeled the degradation of the ligase by heating. We assumed that the ligase degraded exponentially with a time constant of $\tau$ since fresh ligase is introduced by the serial dilution, in the $n$-th round of serial dilution, we assumed $f_n(t) = (1 - d_0)e^{-(t-(n-1)t_0)/\tau} + d_0 f_{n-1}(t)$ with $f_1(t) = e^{-t/\tau}$ where $d_0 = 1/6$ is the dilution ratio and $t_0$ = 50 cycles the period between serial dilutions. A degradation time scale of $\tau$ = 80 cycles reproduced the experimental data well and could be expected even for a thermostable DNA ligase under our thermal cycling conditions. No additional modeling was needed to account for the binding of degraded ligase, suggesting that the degraded ligase does not actively block other ligation reactions.

The serial dilution in the simulation followed the experimental protocol: after 50 cycles, all strand concentrations were diluted by 1/6 except the 20mer strands which were also fed with [X] → 1/6 [X] + 5/6 $X_0$, with $X_0$ = 33 nM the feeding concentration. Only for Figure 7 and Figure S8, a continuous degradation rate of $3.6\,\%/\text{cycle} = -(1/50)\ln(1/6)$ was used for simplicity. To illustrate how above building blocks converge into a system of rate equations, we show the simplest example, the reaction of Figure 4 for the cooperative case of AB and BC. Above rules lead to the rate equations (9). The effective ligation rate is $\hat{k} \equiv k \exp(-t/80) / K_{D,20}^2$ with an affinity ratio $p \equiv K_{D,20}/K_{D,40} - 1$.

$$[\dot{A}] = -\hat{k}([A][B]([AB] + [ABC]) + [A][BC]([AB] + (1 + p)[ABC]))$$

$$[\dot{B}] = -\hat{k}([A][B]([AB] + [ABC]) + [B][C]([BC] + [ABC]))$$

$$[\dot{C}] = -\hat{k}([B][C]([BC] + [ABC]) + [AB][C]([BC] + (1 + p)[ABC])) \tag{9}$$

$$[\dot{AB}] = \hat{k}([A][B]([AB] + [ABC]) - [AB][C][BC] - (1 + p)[AB][C][ABC])$$

$$[\dot{BC}] = \hat{k}([B][C]([BC] + [ABC]) - [A][BC][AB] - (1 + p)[A][BC][ABC])$$

$$[\dot{ABC}] = \hat{k}([A][BC][AB] + [AB][C][BC] + (1 + p)[A][BC][ABC] + (1 + p)[AB][C][ABC])$$

***Long term cooperative replication with serial dilution.***

To test how the cooperative ligation affects the survival of sequences, we performed long term replication experiments with various combinations of templates that were serially diluted. Every 50 cycles, 1/6th of the volume of the solution was diluted with a fresh solution containing 20-base DNA substrates and Taq DNA Ligase. The dilution effect allowed the simulation of an exponential long-term degradation of the template sequence and amounted to a simulated degradation of 3.6 % per cycle against which the replication has to counteract to maintain the initial sequence pool of sequences.

First, the non-cooperating template pairs BA and BC were investigated at an initial concentration of 0.01 nM (Figure 5a). Both withstood the exponential degradation, demonstrating their exponential replication, and settled into a steady state determined by the rates of replication and serial dilution. This mutual symmetry broke down as soon as the motif CA was initially present (Figure 5b). The sequence pattern ‹BA› was suppressed and approached extinction, but the sequence motif <BC> survived together with <CA>. It has to be noted that the need for substrates was identical for both cases: all motifs compete with a second one for the substrates A, B or C.

What broke the symmetry in favor for <BC> and against <BA>? Again, three-letter sequences emerged from two-letter templates. The meta-sequence <BCA> assembled from <BC> and <CA> and helped both <BC> and <CA> in their replication (Figure 4b) but offered no binding site for templated ligation with a cooperative partner for <BA>. The alternative system where initially the template AC was added instead of <CA> inverted the preference and <BC> died out. This showed that above asymmetry was not due to a thermodynamic bias of one particular 20mer substrate or from a sequence dependent ligation rate since now the cooperating 60mer sequence BAC emerged (Figure 5c). The simulation, based on the parameters found from Figure 2, confirmed the nonlinear growth and provided a quantitative description for the experiment (Figure 5a-c, solid lines). The length dependence of competitive ligation offered an enhanced replication of long consensus sequences and tipped the balance towards the sequences which could collaborate by templated ligation with already existing motifs.

### *Kinetics of cooperation can overcome a thermodynamic bias*.

So far, sequences were chosen with similar hybridization strength for the same length of the oligonucleotide sequence. This poses the question as to how a concentration bias could outcompete over sequences which had a stronger hybridization and therefore a faster ligation kinetics in a frequency dependent manner of Figure 5. If this would not have been the case, stronger binding sequences would still invade even the majority of the population. To test this possibility, a sequence bias "*s*" was introduced to the ligation reactions by enhancing the thermodynamic binding stability of A and B compared to that of C. To keep the change symmetric, the $K_{D,20}$ for A and B was reduced to $K_{D,20}/\sqrt{s}$ and increased $K_{D,20}$ for C to $K_{D,20}\sqrt{s}$ (Figure 5d). For consistency, we modified the dissociation constants of longer sequences and used $K_{D,40}/s, K_{D,40}, K_{D,40}, K_{D,60}/\sqrt{s}$ for BA, BC, CA, and BCA, respectively.

We started a simulation with the initial concentrations of templates BA, BC, and CA as in the experiment of Figure 5b. We plotted the concentration ratio of <BC> over <BA> after 1000 temperature cycles versus the binding ratio of $K_{D,20}$ over $K_{D,40}$ which determined the cooperativity of the ligation network. The ratio indicated how much the initial concentration could determine the final replication outcome. For a vanishing length dependence of ligation i.e. by switching off the advantage of cooperative ligation with $K_{D,20} / K_{D,40} = 1$, <BA> dominated over <BC> due to the introduced thermodynamic bias of binding. For example, for *s* = 1.3, as $K_{D,20} / K_{D,40}$ approached 10, the cooperativity could overcome the thermodynamics bias. The motif <BC> now dominated over <BA> due to its faster ligation kinetics, a situation also formed at the experimental value of $K_{D,20} / K_{D,40} = 43$. Based on the model, cooperative ligation could therefore overcome a significant thermodynamic binding bias. This allowed the system to amplify sequence motifs once they are present at an initially higher concentration even if they bind with lower affinity. The result would be an enhanced diversity of the accessible sequence space for evolution.

### *Frequency dependent selection.*

In the previous experiments of Figure 5, the length of the ligating sequences was limited to 60mers due to the limited initial template set lacking templates that form overhangs necessary for an elongation. The same selection of the majority sequences was found for fully ligating templates. Now the sequences grew to considerable lengths (Figure 6b). We compared the replication of two competing groups of cooperating templates. On the one hand the templates AB, BC, CA supported the common, periodic motif ...ABCABC.... When starting with templates CB, BA, AC, the reverse motif ...CBACBA... would be expected to emerge. Those two motifs are two out of six possibilities of the two-letter sequences to cooperate (Figure 7 and supplement S6). Each of them is not promoting the ligation of the other two-letter sequences.

As seen in Figure 6a, the sequences which initially had a majority concentration established and survived in a steady state, while the minority sequences decayed exponentially seen already in Figure 5. For both opposing biases, we observed splitting of the growth kinetics and confirmed that the initial concentration bias was amplified, even though the replication at lower concentrations was faster due to reduced saturation effects. The simulations of cooperative ligation predicted the selection dynamics. In contrast, an exponential replicator without interactions would have immediately replicated all sequences to the identical, high concentration level as documented in supplement S7.

As seen in Figure 6b, the length distribution was initially exponential. But after several cycles, long strands accumulated and formed non-exponential fat tails. Similar shaped distributions were predicted by ligation models previously[16,31,32]. In our experiments, we found sequences with more than 160 bases, offering a good support for the subsequent concentration dependent selection of even more complex sequences.

### *Spontaneous symmetry breaking in simulated cooperative ligation networks.*

As we validated the theoretical model with the experiments, we used the model to extrapolate how individual cooperating sequences emerged locally. Interestingly, the formed sequence patterns persisted against the mixing mechanism of diffusion. Due to the nonlinear growth kinetics, the initial state with a near-uniform sequence concentration was found to be inherently unstable.

To extrapolate how the replication dynamics would amplify small concentration fluctuations, we performed a long-term simulation (Figure 7). In addition to assuming a well-mixed situation (Figure 7a), we implemented molecular diffusion along one dimension. The experimental implementation of feeding, dilution flows, and diffusion in physical space of such a setting has been realized with microfluidics for the autocatalytic replication for the case of CATCH and Qβ replicase[27,33]. However, in the given setting a similar stable long-term observation of ligation under thermal cycling and with the appropriate spatially resolved sequence analysis would be a highly challenging experiment.

The initial concentration of all two-letter template sequences (AA, AB, AC, BA, …) was superimposed with 5% random concentration fluctuations. Despite molecular diffusion, sequence patterns emerged after many temperature cycles. The combination of surviving two-letter sequences can be understood from the hierarchical replication structure. For example, AB, BC, CA cooperated towards sequences ...BCABCA... (Figure 7b, red) whereas BC, CB, AA converged to ...BCBC... and ...AAAA... (Figure 7b, yellow), representing two examples of 6 possible sequence patterns the system can converge to (supplementary Figure S8).

The simulation parameters translated to a 5 mm wide experimental system when using typical diffusion coefficients of the simulated oligomers[34] $D = 10^{-9}$ cm$^2$/s and a temperature cycle of 30 s. The control simulation without cooperative rates of ligation ($K_{D,20}$ / $K_{D,40}$ = 1) produced no patterns (Figure 7c). Also, systems under well-mixed conditions but an initial 5% concentration fluctuation converged towards a randomly chosen single cooperative network where all other competing sequence networks died out (supplement S6). This demonstrated in simulation that the nonlinear selection by the cooperative and hierarchical replication could amplify small fluctuations and spontaneously break the symmetry in sequence space.

### *Discussion*

We argue with our experiments that cooperative interactions between complementary single stranded DNA sequences under templated ligation created replication networks that implemented two important properties of the hypercycle hypothesis[20,25,35]: (i) the achievement of a faster than exponential growth that resulted in a frequency-dependent replication, stabilizing the replication of past majorities sequences and (ii) the inherent selection dynamics for the cooperating sequences in the ligation network. Both dynamic features were not due to special catalytic sequences at or near the ligation site[36,37], but caused by the simple fact that sequences that elongated by ligation were able to bind stronger and therefore ligated faster. The mechanism only required the minimal replication chemistry of ligation.

In the past, a number of theoretical explorations have indicated an interesting dynamic of ligation using coarse grained modeling[31,38-41]. However the lack of kinetic and thermodynamic details and the frequent inclusion of experimentally not supported catalytic function prevented these models to yield quantitative experimental predictions for the presented cooperative ligation. The model by Tkachenko and Maslov indicated the possibility of nonlinear growth for templated ligation. To our knowledge, no experimental demonstration of nonlinear growth for cooperative ligation has been shown. Other modeling work on reaction networks[42-44] did not take into account sufficient mechanistic details to allow direct experimental implementation.

With the higher-order growth modes, the concentration of templates could enhance the speed of replication. The experiments showed how majority sequence networks could suppress minority sequence networks. Growth became not only a function of the ligation rate. A network of cooperating ligation reactions could compensate for weaker binding (Figure 5d). The majority sequence could survive despite the sub-optimal replication rate from inferior binding. The sequence history of replication became important: once a cooperating majority sequence pattern emerged, it was inherently more stable and could defend itself favorably against emerging mutations, stabilizing the initially emerging sequences from statistical fluctuations. If we could include sequence degradation by hydrolysis into the reaction, we expected that the additionally provided sequences for templating would further stabilize the sequence majority.

This dynamic was seen in experiment. High initial concentrations triggered a dominant ligation network that emerged from the cooperative hybridization dynamics between the sequences in the initial pool[45]. The nonlinear enhancement of growth by sequence matching was observed in experiment already for the simplest binary cooperation (Figure 4) but was also in trimeric cooperation networks (Figure 5b,c) and for the formation of long sequences under long term ligation (Figure 6a). The theoretical modeling supported the idea that the cooperativity was caused by feedback and feed forward mechanisms. For example, in Figure 4, sequence matching 40mer two-letter sequences concatenated into a feed forward direction to form 60mers with a three-letter sequence. In addition, a feedback loop back to two-letter sequences was created: the three-letter 60mers templated and replicated an increased amount of two-letter sequences from single letter sequences. The theoretical analysis found that the cooperation strength is given by the binding enhancement $p = K_{D,20} / K_{D,40} - 1$ (eq. 9). Nonlinear growth and selection were observed only for $p > 0$ (Figure 8). The value of $p$ was determined experimentally to about 40 (Figure 2).

In the mechanism, the whole sequence determined the template binding for ligation. Specific bases at the ligation site were not essential for the mechanism. This contrasted with first-order base-by-base modes of replication[13,21], where sequences patterns with the highest thermodynamic stability have the tendency to dominate, leading to sequences biased towards G and C and presumably giving rise to a low sequence complexity. But molecules with desirable catalytic function such as translation require sophisticated higher order structures, which would be unlikely to develop from a purely thermodynamically dominated replication. In contrast, frequency dependent selection from higher order growth opened new routes for the emergence of complex functional molecules.

To study the basic properties of the cooperative cascade networks, we limited the number of unit blocks to three (A, B, and C). However, since the derivation of the nonlinear growth equations (12) did not explicitly depend on the number of unit blocks, the nonlinearity and the symmetry breaking shown in the result section should have also taken place with more unit blocks. We, indeed, expect that the use of more unit blocks would provide a richer selection dynamic. However, at the same time, the increase in the number of units would have reduced the probability of binding between complementary DNA strands. While the probability to bind a complementary sequence would drop, it is not obvious why the concentration window for the onset of nonlinear growth would have changed with respect to the concentration regime of ligation.

Our proposed mechanism considerably differed from second-order growth models such as the hypercycle. In these models, the replication rate became proportional to the number of replicates, leading to a hyperbolic growth (eq. 13). Unfortunately, this also implied that for degradation conditions, mimicked in our experimental setting by serial dilution, the initial strand concentration must be kept over a concentration threshold. Otherwise the replicators would vanish due to the lack of a fast replication kinetics (Fig. 8c). In contrast, the nonlinearity of a cooperative cascade of ligation offered a non-vanishing replication rate for low replicate concentrations. Even vanishing concentrations can be replicated in principle (eq. 12, Fig. 8b). We did not clearly observe a downward convexity in the growth curves (Fig. 4, 5, and 6). While the hypercycle had a strong downward convexity because of its pure second-order nature (Fig. 8c), the convexity is subtle in the present system because the first-order and second-order reactions are superposed (Fig. 8b). A significant improvement in the quantification would be necessary to detect the subtle convexity.

It has been argued that classical hypercycles were sensitive to parasites that could take a free ride on the catalytic functions of the cycle[46,47]. The reason was that the cooperation is asymmetric and the product of one cycle acted as the replicating machinery for the next cycle. In the shown cooperative ligation networks, however, the catalytic role played by the template is symmetric. If there is a reaction where a strand X works as a template to produce a strand Y, there is also the reverse reaction where Y acts as a template to produce X. Therefore, asymmetric free riders of one sequence which would not also offer themselves as template would be difficult to imagine. The symmetric nature of the ligation network, in contrast to hypercycles, should therefore be much robust against simple sequence parasites.

In the past, well-balanced combinations of protein-catalyzed DNA reactions using multiple proteins have been shown to implement complex algorithms[26,48,49]. These experiments included higher-order growth dynamics. But these implementations were created from a fine-tuned network of complex protein functions which were hard, if not impossible to imagine prebiotically. In contrast, we used a protein to speed up a prebiotically plausible ligation reaction. The mechanism we studied did not require a particular catalytic enzyme, but the basic combination of hybridization and ligation. This reaction was likely implemented by the prebiotic activation chemistry that would have also triggered the random polymerization of the nucleotides in the first place.

Previously, replication mechanisms have been studied using the complex catalytic function of proteins to implement base-by-base replication such as Qβ replicase[21] or a combination of a reverse transcriptase and a polymerase[29]. These proteins generally showed particular sequence dependencies and non-trivial reaction mechanisms such as template detection. The Qβ replicase under high salt concentrations could lead to an arbitrary elongation of sequences by untemplated bases[50]. However, under kinetic competition, as done in the classical experiments of Spiegelman[21], these replication mechanisms had the tendency to shorten the sequence length by mutations.

In comparison, the templated concatenation by ligation showed the inherent tendency to produce sequences longer than the initial templates. And since longer sequences offered more stable binding sites, they were able to cooperate with other sequences, and longer strands were favored in the ligation networks. We found a fat tail in the length distribution in our experiments (Fig. 6b) which enabled significantly more complex sequences to emerge, enhancing the possibility of finding catalytically functional oligonucleotides. In the past, theoretical and experimental analysis of ligated sequences confirmed a reduction in the sequence diversity.[16,31,32].

We measured the sequence dependence of the ligation kinetics and used it to model the ligation experiments (Figure 2). We found that the ligation rate k was constant and independent of the strand length once a bound template complex was established. However, the binding affinity $K_D$ determined the formation of the templated pre-ligation complex and modulated the effective rate of ligation. Both kinetic rates were kept constant in the modeling.

Only between experimental implementations, the rate *k* only required slight adaptations to fit the experiments, likely caused by saturating the catalytic activity of the protein due to varying concentrations of ligation sites. In addition, different batches of the ligase showed slightly varied thermal degradation. Apart from these minor adaptations of the overall ligation characteristics, no fits of parameters were required to describe the experiments with the theoretical models. This is why we would be confident to use the simulations to predict the system's behavior in scenarios which would have been experimentally too difficult to perform reproducibility.

In this study, we limited the number of unit building blocks to three pairs (A, B, and C) in order to study the basic properties of the cooperative cascade and still be able to follow the dynamics with theory. However, there is no reason to assume why the shown nonlinear replication that enable symmetry breaking would not take place for example for a more diverse sequence population or for shorter sequence lengths at lower temperatures. For example, the derivation of the nonlinear growth equations (12) did not depend on the number of different short sequences. We expect that the use of more starting sequences provided a richer dynamic. But at the same time, the increase in sequence space would reduce the probability for a matching hybridization to happen, resulting in the need to enhance the concentrations of the starting strands in the experiments.

In addition to the numerical simulations, the analytical theory in the appendix was used to describe how the cooperativity caused the instability of the uniform state in the sequence space. The amplification from small concentration perturbations merged into patterns of majority networks and made them stable in sequence space. The amplification of the fluctuations was predicted to produce different dominating sequence patterns at different locations even under the presence of diffusion. Once the spatial pattern formed, it would have been stabilized by the frequency-dependent selection, suppressing the growth of competing sequences diffusing from neighboring patterns. It is likely that in the long term only one sequence pattern remained. Before, the coexistence of multiple sequence configurations could only be realized without compartments such as membranes, limiting the potential for lateral gene transfer. It should be noted that the spatial pattern formation based on hypercycle, CATCH, and Qβ replicase characteristics has been explored previously by simulations[26,46] and in impressive microfluidic experiments[26,32,50]. An experimental test of the spatial coexistence of different sequence information under ligation and thermal cycles is one of the next experimental challenges.

In this study, ligation provided a primitive mode for the replication for sequences. As shown, the inherent cooperative characteristic of ligation would likely stabilize the replication of sequence information. It should be noted that for the full exploration of long sequences, the thermal cycling seems important to shuffle sequences between templates and to enable exponential replication. Limited versions of the dynamics should be however already seen for short sequences where the off-rate is high enough to allow spontaneous exponential ligation chain reactions.

To implement temperature cycles, a heat flow, for example across elongated rock pore systems on an early Earth[34,52], could provide thermal cycling by microfluidic convection[53]. This could also provide the enhanced concentrations of oligonucleotides for templating and possibly increase their polymerization dynamics[18]. Combined with a flow-through, the thermal gradient could localize replicated oligonucleotides in a length selective manner and supply the replication reactions continuously[24]. It should be noted that the shown mechanism bears a strong similarity to chemical systems that were able to amplify a bias between chiral molecules[54]. We therefore hypothesize that the concentration dependence of the ligation network could also purify backbone heterogeneity based on their differential duplex stability[55].

Ligation was a comparably simple reaction and the chemical details of its prebiotic implementation have been explored[10,56-60]. It will be interesting to see how similar cooperative mechanisms could be established in base-by-base replication mechanisms. Replication of RNA or DNA from single bases using ribozymes has progressed continuously as shown by Joyce[7] and Holliger[61]. Non-catalytic replication is a very interesting third possibility and recent experiments by Szostak showed fast progress[62], becoming more attractive after better understanding its prebiotic plausibility[17]. In the future, it would be also interesting to integrate the network capabilities of RNA-based recombinases[6,14,63] to explore how the emergence of recombination as a most simple catalytic activity of RNA could implement modes of cooperative replication dynamics under thermal cycling.

## *Conclusion*

Our experiments showed that a most basic mechanism - the joining of two DNA strands on a third template strand in a pool of diverse sequences - formed a cooperative cross-catalytic reaction network with higher-order replication kinetics. The shown process offers a mechanism that implements in a significantly more simple and robust manner the nonlinear replication dynamics first proposed by hypercycles. This nonlinear ligation chain reaction could circumvent central drawbacks of hypercycles, namely the concentration growth threshold under molecule degradation (Fig. 8c) and the instability against 'viral' molecules using the replication dynamics, but not contributing to it.

It remains to be seen how the mechanism could amplify a majority sequence bias from more than the shown six sequences, i.e. from a more diverse random sequence pool. We expect that the hybridization dynamics of 20mers would offer much more sequences to implement the shown mechanism. Unfortunately, it is challenging to extrapolate the brute force theoretical modeling to more complex experiments. In the future, we expect long term experiments using deep sequencing to offer more insights into this highly nonlinear, but realistic replication dynamics which was possible when the first oligonucleotides emerged.

## *Overview over Supplemental Material.*

The supplemental material has additional details on both the experiments and the theoretical modeling. It is structured as follows: **S1. Ligation reactions.** Experimental protocols, sequences and melting temperature measurements. **S2. Quantitative PCR.** Protocol and calibration of 40mer two letter quantification with cold PCR. **S3. Dissociation constant and ligation rate.** Kinetic experiments to determine the ligation rate. **S4. Gel electrophoresis.** Determination of the length distribution with denaturing electrophoresis. **S5. Numerical simulations.** Strategy, details and examples of the numerical calculation. **S6. Stochastic emergence of self-sustaining sequence structure.** Documentation of numerical calculations demonstrating the symmetry breaking in bulk solutions. **S7. Theoretical analysis.** One dimensional analytical models of competitive higher-order growth. **S8. Supplementary experimental data.** Documentation of replicates of the experimental results.

## *Appendix A: Nonlinear growth mechanism.*

We analyzed an analytical model to elaborate on the mechanism of the higher-order growth and the frequency-dependent selection. Instead of the huge set of equations used in the numerical simulations (supplement S5), we used simplified rate equations. We consider a simple cooperative system starting with the templates AB and BC. This corresponds to the experiment shown in Fig. 3 but here we added dilution and feeding kinetics. The rate equations are

$$[\dot{A}] = -\hat{k}\big([A][B]([AB]+[ABC]) + [A][BC]([AB]+(1+p)[ABC])\big) - d([A]-S)$$

$$[\dot{B}] = -\hat{k}\big([A][B]([AB]+[ABC]) + [B][C]([BC]+[ABC])\big) - d([B]-S)$$

$$[\dot{C}] = -\hat{k}\big([B][C]([BC]+[ABC]) + [AB][C]([BC]+(1+p)[ABC])\big) - d([C]-S) \quad (10)$$

$$[\dot{AB}] = \hat{k}([A][B]([AB]+[ABC]) - [AB][C][BC] - (1+p)[AB][C][ABC]) - d[AB]$$

$$[\dot{BC}] = \hat{k}([B][C]([BC]+[ABC]) - [A][BC][AB] - (1+p)[A][BC][ABC]) - d[BC]$$

$$[\dot{ABC}] = \hat{k}([A][BC][AB] + [AB][C][BC] + (1+p)[A][BC][ABC] + (1+p)[AB][C][ABC]) - d[ABC]$$

Here, $\hat{k} \equiv k/K_{D,20}^2$ and $p \equiv K_{D,20}/K_{D,40-1} - 1$. $S$ is the amount of the substrate supply, $d$ is a degradation rate. The growth curves obtained by simulating (10) with $p$ = 50 and $p$ = 0 are plotted in Fig. 8a. We see an enhanced growth of ABC with $p$ = 50 which eventually dominated the <AB> and <BC> motif concentrations. After AB and BC accumulated, ABC grew due to the faster ligation rate (Fig. 2). This demonstrated the positive feedback mechanism by the cooperation of AB and BC. ABC mediated the cooperation between AB and BC and accelerated the growth of <AB> and <BC> with a downward convex. Without the enhanced ligation speed by a longer overlapping ($p$ = 0), we did not observe the higher-order growth and find a lower steady state concentration of <AB> and <BC>. These plots show that two factors account for the higher-order growth: the cooperation mediated by an elongation and the enhanced ligation rate for longer strands.

The growth curves of the cooperative cascade model (Fig. 8b) and the simplest hypercycle model (Fig. 8c) calculated starting from varied initial concentrations clearly showed the difference of the two models. Both showed a nonlinear growth, but with different characteristics. In the hypercycle model (Fig. 8c), the initial growth rate decreased as the initial concentrations decreased. A growth overcoming dilution is observed only with an initial concentration above some threshold. On the other hand, in the cooperative cascade model, the initial growth rates were similar for all initial concentrations. When the concentrations exceeded a threshold, concentration indicated by a dotted horizontal line, the reaction was accelerated. The threshold is given by approximately 2[A] / $p$, which is indicated by a dotted horizontal line in Fig. 8,b as we will see.

To understand these behaviors, we simplified the rate equations. By rearranging (10), we obtain the rate equations of the sequence motifs $\langle AB \rangle \equiv [AB] + [ABC]$ and $\langle BC \rangle \equiv [BC] + [ABC]$:

$$\langle \dot{AB} \rangle = k'[A]\big((S - \langle AB \rangle)\langle AB \rangle + p[BC][ABC]\big) - d\langle AB \rangle,$$
$$\langle \dot{BC} \rangle = k'[C]\big((S - \langle BC \rangle)\langle BC \rangle + p[AB][ABC]\big) - d\langle BC \rangle. \quad (11)$$

By reasonably assuming the symmetry [A] = [C] and [AB] = [BC], we obtained the rate equation for $\alpha \equiv$ <AB> = <BC> according to:

$$\dot{\alpha} = k'(S-\alpha)^2(1 + pL_\alpha)\alpha - d\alpha. \quad (12)$$

Here, $L_\alpha \equiv ([A]/[AB] + [A]/[ABC])^{-1}$ is the harmonic average of [AB] / [A] and [ABC] / [A] divided by 2 which is close to the minimum of either [AB] / [A] or [ABC] / [A]. Because $L_\alpha$ quantifies the accumulation of long strands, the reaction rate $\tilde{k} = k'(S-\alpha)^2(1 + pL_\alpha)$ increases as the reactions proceed and the long strands accumulate. The growth is enhanced when the nonlinear term $pL_\alpha$ exceeds one. This threshold corresponds to [ABC] ≈ [A] / $p$ or approximately $\alpha$ ≈ 2[A] / $p$ in the present setup.

The hypercycle model based on the template-directed ligations with unknown hypothetical catalytic function by oligonucleotides is theoretically studied by Wills *et al.* [40]. They assumed an unknown hypothetical catalytic ability by oligonucleotide to construct a hypercycle model. The rate equations of the hypercycle is reduced to

$$\dot{x} = k'(S - x)^2 x^2 - dx. \tag{13}$$

These simplified rate equations (12) and (13) explain the growth modes seen in Fig. 8b and c. The hypercycle (13) has a strong nonlinearity with the reaction rate proportional to *x*, which vanishes as *x* goes to zero. Therefore, a growth that can overcome the serial dilution is not observed when started from low initial concentrations. On the other hand, in the cooperative ligation cascade model (12), the reaction rate does not vanish even at a vanishing concentration $\alpha$ because of the moderate nonlinearity $(1 + pL_\alpha) \alpha$. Therefore, if the reaction rate $\tilde{k}$ is larger dilution rate *d*, positive growth is always observed independent of the initial concentrations. The system grows in a manner that the reaction rate $\tilde{k}$ increases gradually from a slow rate $k'(S - \alpha)^2$ to a faster rate $k'(S - \alpha)^2 (1 + pL_\alpha)$ as the concentration $\alpha$ increased.

### *Appendix B: Concentration Dependent Selection of sequence motifs*.

We consider another cooperative sequence group $\beta$, which is, for example, <BC> and <CA> and shares the same substrates A, B, C with $\alpha$. The rate equation for the competitive growth of $\alpha$ and $\beta$ leads to

$$\begin{pmatrix} \dot{\alpha} \\ \dot{\beta} \end{pmatrix} = k'(S - \alpha - \beta)^2 \begin{pmatrix} (1 + pL_\alpha)\alpha \\ (1 + pL_\beta)\beta \end{pmatrix} - d \begin{pmatrix} \alpha \\ \beta \end{pmatrix} \tag{14}$$

The selection dynamics simulated by (14) is shown in Figure 8d. The frequency dependent selection was observed only when *p* > 0, similar to the results in Figure 6a. This behavior can be explained by a linear-stability analysis of (14). The stability of the uniform state, $\alpha_0 = \beta_0 \neq 0$, is determined by the sign of *p* (see supplement S7.2). When *p* > 0, uniform state is unstable, and either $\alpha$ or $\beta$ becomes dominant. When *p* = 0, uniform state is neutral. The initial difference is kept for a long time. This neutral behavior with *p* = 0 was observed because the simple model by (14) neglected the effect of product inhibition, which causes the convergence of $\alpha$ and $\beta$ to the same value.

### *Appendix C: Spatial pattern*.

The pattern formation observed in the simulation (Figure 7b) can be modeled by adding a diffusional term to (14):

$$\frac{\partial}{\partial t} \begin{pmatrix} \alpha(x,t) \\ \beta(x,t) \end{pmatrix} = k'(S - \alpha - \beta)^2 \begin{pmatrix} (1 + pL_\alpha)\alpha \\ (1 + pL_\beta)\beta \end{pmatrix} - d \begin{pmatrix} \alpha \\ \beta \end{pmatrix} + D \frac{\partial^2}{\partial x^2} \begin{pmatrix} \alpha \\ \beta \end{pmatrix} \tag{15}$$

The linear-stability analysis showed that, the uniform state is unstable when *p* > 0 and a spatial pattern forms spontaneously if the wave number is less than $\sqrt{kp\alpha_0(S - 2\alpha_0)/D}$ (supplement S7.2). This inequality suggests that the pattern forms only when *p* > 0, which is consistent with our observation in the simulation (Fig. 7b). The pattern becomes finer with smaller diffusion and larger reaction rate. The spatial instability is caused by the inherent instability of the reaction in the sequence space and is not a spatial effect. The formed spatial patterns are stabilized by the frequency-dependent selection which suppresses the growth of other sequence patterns diffusing from neighboring pattern.

### *Acknowledgments.*

This work was supported by the Alexander von Humboldt Foundation, European Research Council (ERC) Starting Grant AutoEvo and Advanced Grant EvoTrap, the CRC Emergence of Life project P07 and P09, the SFB 1032 Project A04, a grant from the Simons Foundation (SCOL 327125, DB), and JSPS KAKENHI (15H05460). We very much thank Jonathan Liu, Patrick Kudella, Alexandra Kühnlein and Noel Martin for corrections on the manuscript.

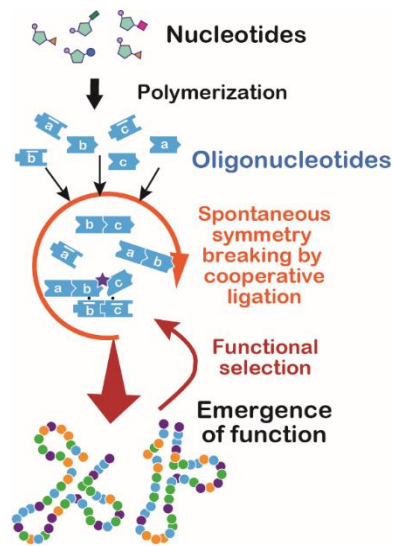

**Figure 1: Potential role of cooperative ligation in early molecular evolution of DNA oligonucleotides.** Schematic representation of a potential path towards the selection and emergence of DNA sequences with incipient catalytic activities. Initially, a random oligonucleotide pool could have been originated from the condensation and polymerization of single nucleotides. The same chemical mechanism that catalyzed the oligomerization of nucleotides could have also carried out the ligation of two adjacent oligonucleotides sequences brought into close proximity by a third complementary sequence. This process is termed templated ligation. The template binding would have required a critical oligonucleotide length at the temperature of polymerization. We used 20mer oligomers at 67 °C to minimize the bias of the ligase protein used for ligation. In a pool of sequences, templated ligation becomes cooperative. Our experiments suggest that cooperative ligation could have led to a spontaneous breaking of sequence space symmetry. Depending on the initial concentration of ligated sequences at a given location, a cooperative set of sequences emerged that stabilized their own replication by ligation. Interestingly, the sequences with highest initial concentration dominated the population, a process requiring a non-linear replication characteristic. At a given location this spontaneous symmetry breaking would generate a homogeneous oligonucleotide sequence pool. If this had an incipient function, the whole molecule population would show a significant evolutionary advantage, outcompeting the neighboring locations and creating a starting condition and stabilization for the emergence of Darwinian evolution.

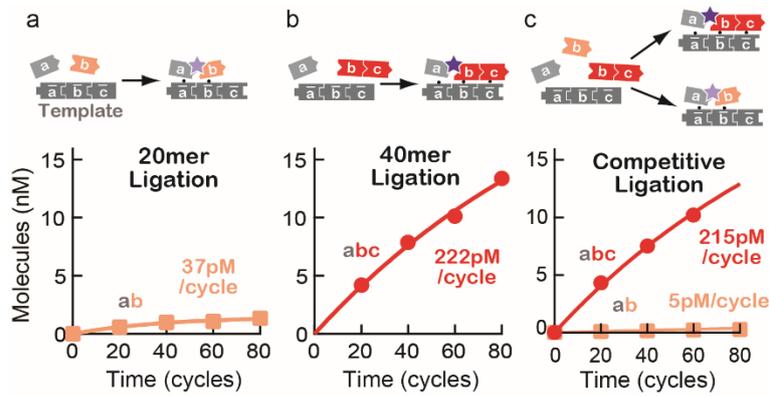

*Figure 2: Competitive ligation kinetics. To model the reaction, the kinetics of ligation was measured for **a)** 20mers **b)** 40mers and **c)** in the competitive case. Two DNA oligonucleotide sequences, **b** and **bc** competed for ligation to a common complementary 60mer template and ligated under thermal cycles between 67 °C for 10 s and 95 °C for 5 s without serial dilutions. The sequences of the 20mers were a: 5'-ATCAGGTGGAAGTGCTGGTT-3', b: 5'-ATGAGGGACAAGGCAACAGT-3' and c: 5'-ATTGGGTCACATCGGAGTCT-3'. Under competitive conditions, the templated ligation that gave rise to **abc** was 40-times faster than the ligation, leading to the **ab** sequence. Without competition, the difference reduced to a factor of 6. The kinetics could be understood by the ratio of the dissociation constants $K_D$ due to the competitive hybridization of **b** and **bc** on the same template sequence (eq. 3). $K_D$ was modelled as an effective dissociation constant under thermal cycling. The $K_D$ for 20-base, 40-base, and 60-base nucleotide sequences could be determined by analyzing the ligation kinetics. The ligation rate k was found to be constant in all three cases.*

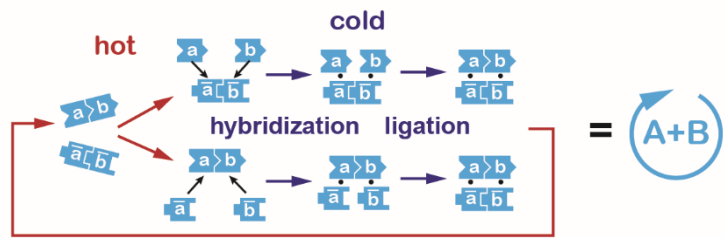
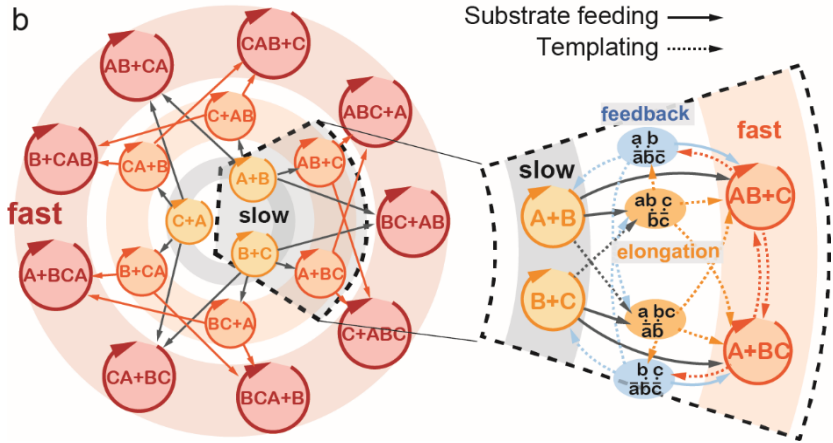

*Figure 3: Cooperative ligation chain reaction.* **a)** *Schematic diagram of a ligation chain reaction depicting the exponential replication of oligonucleotide sequences under thermal cycling. Individual lower-case letters correspond to their 20-base single-stranded DNA sequences.* **b)**, *Schematic diagram of a cooperative ligation cascade network. Templated ligations of partially complementary sequences form a cooperative information-replication network. They are connected by binding to templates (dashed arrows) and by providing substrates (solid arrows). As seen in the zoomed section, the network expands towards longer sequences by ligation-based elongation. Longer sequences have more possibilities to work as templates and substrates and therefore mediate more reactions. Furthermore, the higher stability of longer strands (40-base and 60-base vs. 20-base sequences) lead to faster ligation. This created a nonlinear cascade of cooperative replication reactions with a multistable dynamics. The replication became frequency dependent and sequences with higher initial concentrations were preferred, a characteristic not possible with the simple ligation chain reaction in (**a**).*

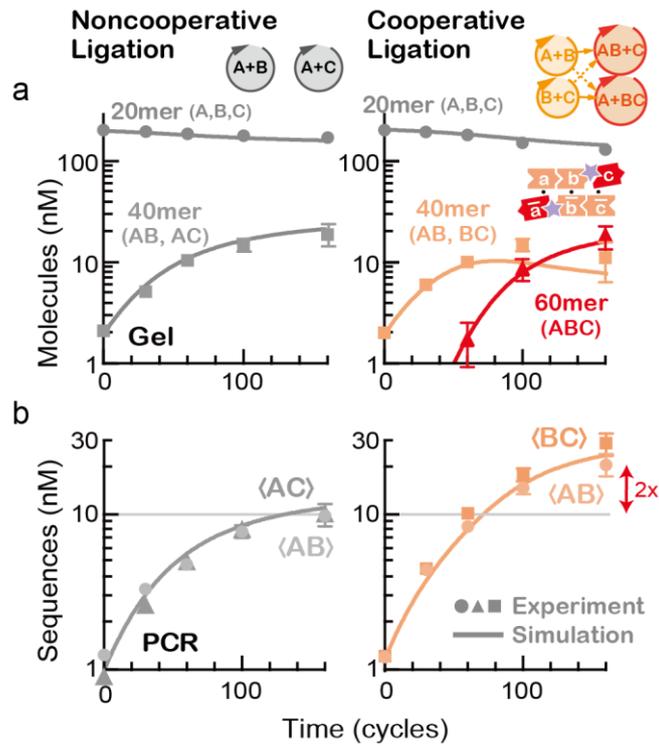

*Figure 4: Enhanced replication by cooperative ligation. a)*, The ligation dynamics of the non-cooperative sequences (AB and AC) was compared with the cooperative sequences (AB and BC). The latter could hybridize at the sequence B, creating a 60-base sequence template ABC for faster ligations of AB and BC. All reactions were supplied with the substrates sequences A = {$a, \bar{a}$}, B = {$b, \bar{b}$} and C = {$c, \bar{c}$}. The concentrations of the oligonucleotides obtained after the ligation experiment were measured by gel electrophoresis as explained in Supplement S4. Solid lines were obtained from simulations (eq. 9) (supplement S5) that considered the experimentally determined $K_D$ and k parameters. *b)*, The concentration of the two-letter 40mer sequence motifs <NN> was quantified by a calibrated real time COLD PCR (supplement S2). The cooperative ligation led to a two-fold faster growth rate of the 40-base sequence two-letter sequence motifs <AB> and <BC> compared to the growth rate of the non-cooperative sequences <AB> and <AC>. For example, <AB> combined the concentrations of the two sequences [AB] + [ABC]. Replicates of the experiment are shown in the supplementary Figure S10

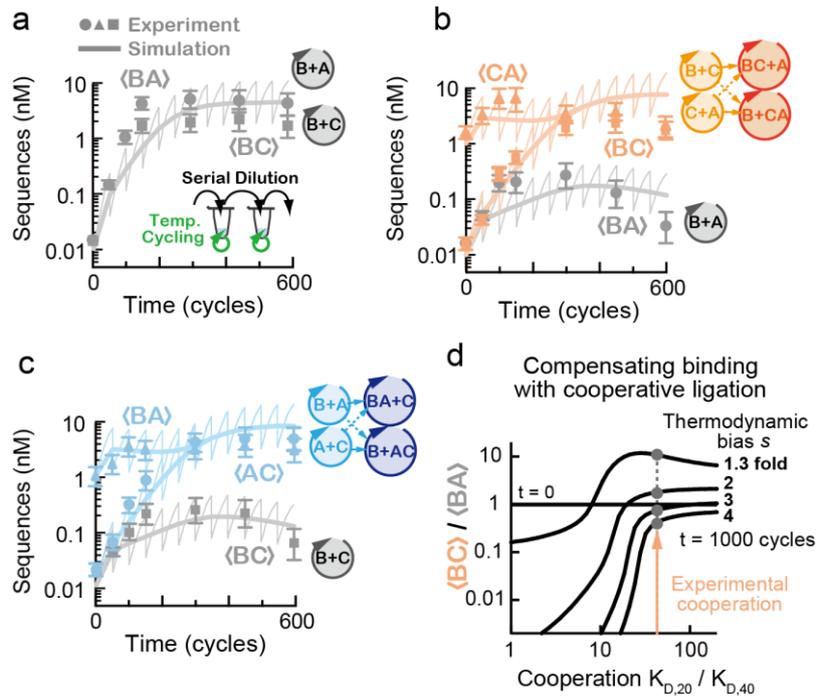

*Figure 5: Concentration dependent replication by cooperative ligation. Cooperative ligation was found to replicate differentially based on the initial concentrations of the cooperating sequences. The reactions were performed under thermal cycling and serial dilution. The dilution simulated molecule degradation and was used to replenish ligase and a complete pool of 20-bases A, B and C, simulating an untemplated polymerization as oligonucleotide source. The three cases were: **a)**, 40-base templates BA and BC were found to replicate from 0.01 nM (initial concentration) to similar steady state concentrations. 60-base sequences could not be formed due to the lack of a common binding motif. **b)**, The same was performed with the addition of 1 nM of CA sequence to the reaction. Since the sequences CA and BC could cooperate by forming the motif BCA, we observed that both the sequences survived the dilution effect. However, the non-cooperating sequence BA died out. The initial concentration bias triggered a symmetry breaking by a concentration-dependent sequence selection. **c)**, Conversely when the BA sequence was provided to the reaction, the system picked AC instead of BC and created the 60mer sequence BAC. The thin lines correspond to mathematical simulations that considered the serial dilution dynamics. Thick curves connect the averages before and after the serial dilutions for comparing the simulation with experimental data. Replicates of the experiment are shown in supplementary Figure S10. **d)**, We tested how the kinetically driven cooperation could compensate a thermodynamic binding bias of the sequences A, B, C. The reaction in (b) was simulated with a thermodynamic bias where sequences A and B bound s-fold better than the sequence C. Without a cooperative ligation mechanism ($K_{D,20} / K_{D,40} = 1$), the sequence BA dominated after 1000 cycles over BC. With the experimental value $K_{D,20} = 43\,K_{D,40}$ the sequence BC dominated over BA. The cooperative kinetics could overcome an up to s=3-fold thermodynamic bias, offering an increased sequence diversity.*

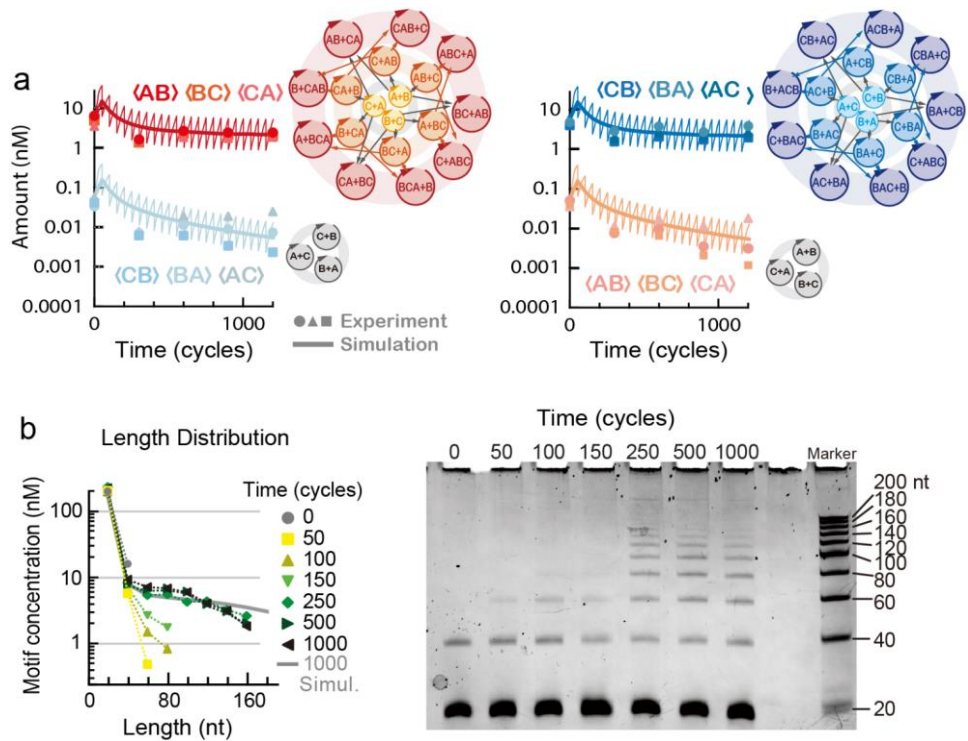

*Figure 6*: **Frequency-dependent selection of cooperative ligation sequences.** *Frequency dependent replication of two competing groups of cooperating ligation patterns. Two groups of three cooperative sequences AB, BC, CA could generate the sequence "...BCABC..." by elongation whereas the sequences CB, BA, AC could cooperate to generate the sequence "...BACBA...". **a)**, The reaction was either started with a concentration bias of 6 nM vs. 60 pM for AB, BC, CA over CB, BA, AC or in the inverted concentration situation. As before, the replication reactions were supplemented with A, B and C sequences and subjected to a serial dilution that mimicked a worst-case molecule degradation without sequence recycling. In both cases, the majority sequence motif patterns were sustained by replication against the serial dilution. The minority sequences died out. The simulation predicted the experimental finding. See supplementary Figure S10 for replicates of the experiment. **b)**, The cooperation generated long oligomers up to 160 bases in length from the initially 40 bases long starting sequences after more than 500 cycles. The length distribution showed fat tails as measured by gel electrophoresis. The simulation predicted this length distribution (solid line).*

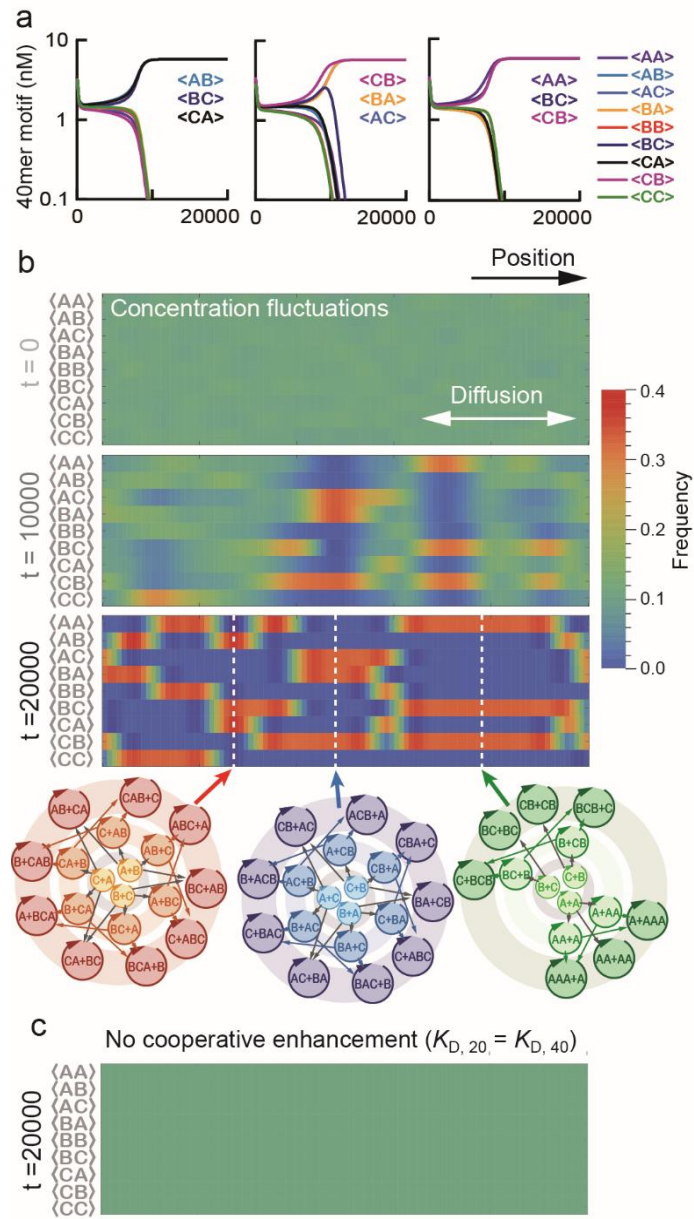

*Figure 7: Spontaneous symmetry breaking. a),* In a simulation of cooperative ligation, sequence structures emerged stochastically under well-mixed situations when we initiated the reactions by a 5% concentration fluctuation in an otherwise uniform concentration of all the nine possible dimers (AA, AB, …, CC). We observed a final steady state dominated by one of six possible cooperative combinations. For example, AB and AC did not coexist because AB and AC were competitive as B and C competed for the right-hand side of A. The dilution rate was d = 0.045 / cycle. *b),* Spatial sequence patterns emerged and coexisted despite of the mixing by diffusion. Cooperating sequences emerged as local patterns after more than 10000 cycles. Examples of the dominant ligation network were shown at the position of the broken lines. They correspond to the sequences in (a). This demonstrates the emergence of individual sequence patterns without the need of compartmentalization. *c),* Control simulation without cooperative enhancement of the ligation from binding ($K_{D, 20}$ = $K_{D, 40}$ = $K_{D, 60}$ = 150 nM). The initial concentration fluctuations disappeared and the system converged to a uniform state.

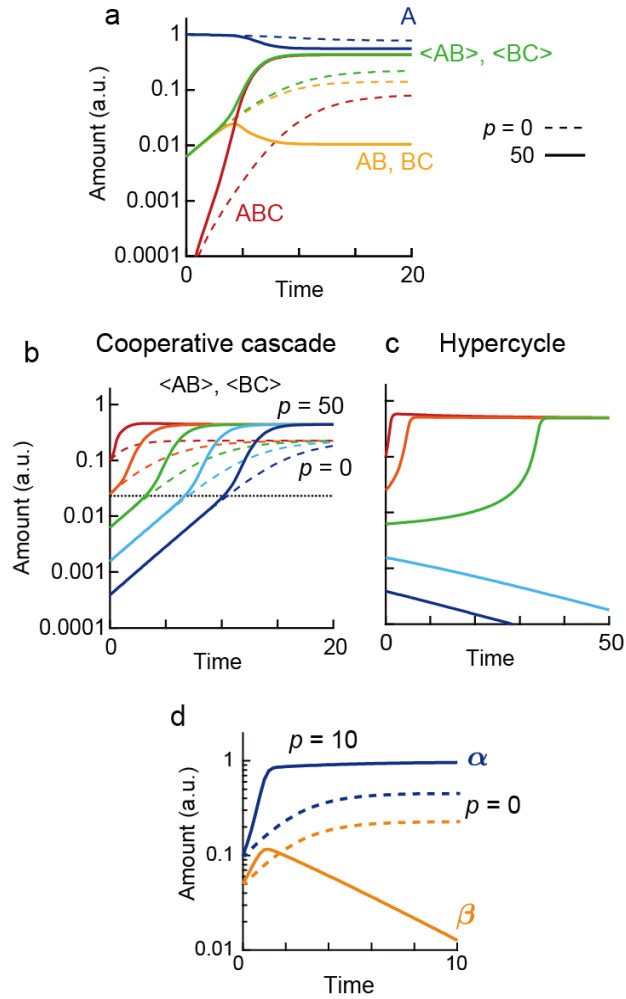

***Figure 8: Ligation cascade in simplified theoretical models.*** *We analyzed the theoretical model of our experimental system. A comparison with a hypercycle model illustrated the difference in their reaction orders and dynamics. **a)**, Cooperative ligation of Figure 4 was simulated with a simplified model of equation (10) with the parameters of effective reaction rate $\hat{k} = 1$, dilution rate d = 0.6, concentration of fed substrate S = 1, and a ligation enhancement by cooperation of p = 50 (solid) or p = 0 (dashed). The initial conditions were A(t = 0) = B(t = 0) = C(t = 0) = 1. ABC(t = 0) = 0. AB(t = 0) = BC(t = 0) = 0.1 / 16. Growth curves of A (navy), AB, BC (orange), ABC (red), and <AB> = AB + ABC, <BC> = BC + ABC (green) with positive cooperativity (p = 50) are shown. **b)**, Growth curves started from different initial concentrations (0.1, 0.1/4, 0.1/16, 0.1/64, and 0.1/256) for the cooperative cascade model (10) with p = 50 (solid) or p = 0 (dashed). The initial concentrations were changed to illustrate the difference in the growth dynamics. The growth is enhanced when the motif amount exceeds a threshold given by <AB> = <BC> ≈ 2[A] / p indicated by a horizontal dotted line. We used the steady state value of [A]. Hypercycles died out against dilution if started at a too low initial concentration. **c)**, Growth curves of the hypercycle model[13,41] with the same initial concentrations as in (**b**). **d)**, The concentration dependent selection of cooperating sequences for the cooperative cascade model with p = 10 (solid) or 0 (dashed). The equations (14) were numerically solved with the parameters k = 1, S = 1, d = 0.5, α(t = 0) = 0.1 and β(t = 0) = 0.05. L$_α$ was approximated by α for simplicity.*

# Cooperative ligation breaks sequence symmetry
# and stabilizes early molecular evolution

Shoichi Toyabe[1, 2] and Dieter Braun[1]

1 Systems Biophysics, Physics Department, NanoSystems Initiative Munich and Center for Nanoscience, Ludwig-Maximilians-Universität München, Amalienstrasse 54, 80799 München, Germany;
2 Department of Applied Physics, Graduate School of Engineering, Tohoku University, Aramaki-aza Aoba 6-6-05, Aoba-ku, Sendai 980-8579, Japan

## S1. Ligation reactions

20-base and 40-base single-stranded DNA oligonucleotides were purchased from Biomers (Ulm, Germany) with PAGE (Polyacrylamide Gel Electrophoresis) or HPLC (High Performance Liquid Chromatography) purification grade. Three complementary pairs of 20-base DNA oligonucleotides ($a$, $\bar{a}$), ($b$, $\bar{b}$), and ($c$, $\bar{c}$) were the minimal DNA sequence units used in this study. Sequences are reported in Table S1. Note that each oligonucleotide contained one-base 5'-overhang to reduce nonspecific ligations[S1, S2]. The melting curves of the 20-base oligonucleotides strands were calculated by deriving the fluorescence intensity curve arising from a solution containing 250 nM of each DNA strands, the Taq DNA ligase buffer and 1x SYBR Green I, a dye that specifically fluoresce upon binding to DNA double strands (Figure S2). The melting temperature ($T_m$) corresponded to the maximum value of the derivative of the fluorescent intensity curve. $T_m$ of the three complementary DNA strand pairs were 67.5 °C.

A typical ligation reaction (Figure S1) consisted of 5 µl reaction mixture containing 0.16 units/µl thermostable Taq DNA Ligase, 1x Taq DNA Ligase Buffer (New England Biolabs, MA, USA) and 33 nM of each of the DNA sequence units. The temperature cycling was set at 67 °C for 10 s and 95 °C for 5 s (Table S2) and was conducted with a real-time PCR cycler (Bio-rad, USA).

Serial dilutions were performed by hand pipetting. Every 50 cycles, one volume fraction of the ligation reaction solution was diluted with 6 volumes of fresh solution containingDNA oligonucleotides (substrates) and the Taq DNA ligase. The dilution ratio was per cycle was ($d$ = 0.036/ cycle).

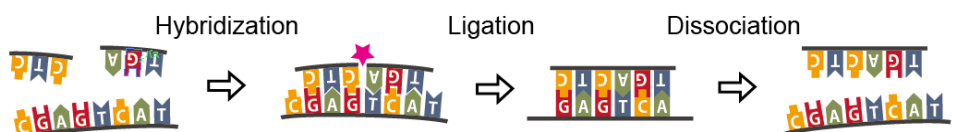

*Figure S1: Schematic representation of the templated ligation reaction mechanism. Complementary hybridization of two short DNA oligonucleotide strands on a longer template strand favors the ligation of the shorter strands. The ligation rate depends on two factors, the DNA ligase dynamics and the hybridization equilibrium constants of the DNA strands. This reaction is fasten by a ligase enzyme. The DNA hybridization is stable when the temperature is lower than the melting temperature ($T_m$) between the two strands. $T_m$ increases with the hybridization overlapping length. Therefore, strand dissociation is typically slow but can be enhanced by raising the temperature. Temperature cycling allows molecules to periodically oscillate between hybridized and dissociated states enabling DNA strands to rearrange, triggering for example the elongation of the DNA strands by sequential ligations. The process is exponential, as the ligase chain reaction[S1].*

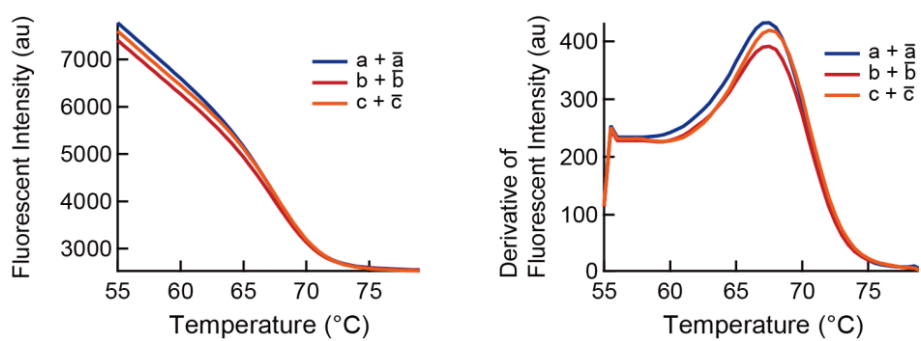

*Figure S2: Calculation of the melting temperature of the 20-base DNA strands.* 250 nM of each DNA oligonucleotides were mixed with the Taq DNA ligase buffer and 1x SYBR Green I. The melting temperature ($T_m$) corresponded to the maximum value of the derivative of the fluorescent intensity curve. $T_m$ of the three complementary DNA strand pairs were 67.5 °C.

*Table S1*: List of the 20-base DNA sequences used in this study consisting of three pairs of complementary sequences. Highlighted in bold are the one-base 5'-overhang.

|   | Sequences |
|---|---|
| A | $a$:  5'- **A**TCAGGTGGAAGTGCTGGTT  - 3'<br>$\bar{a}$: 3'-   AGTCCACCTTCACGACCAA**T** - 5' |
| B | $b$:  5'- **A**TGAGGGACAAGGCAACAGT  - 3'<br>$\bar{b}$: 3'-   ACTCCCTGTTCCGTTGTCA**T** - 5' |
| C | $c$:  5'- **A**TTGGGTCACATCGGAGTCT  - 3'<br>$\bar{c}$: 3'-   AACCCAGTGTAGCCTCAGA**T** - 5 |

*Table S2: Ligation reaction protocol.* The temperature cycling protocol used for the ligation reaction experiments consisted of an initial denaturation step at 95 ºC for 20 s followed by 40 or 50 cycles where for each cycle double stranded DNA sequences were denaturated at 95 ºC for 5 s and subsequently annealed and ligated at 67 ºC. After 40 or 50 cycles, the samples werethen heated at 80 °C for 1 s, taken from the PCR cycler, cooled down on ice, and diluted with a fresh solution containing Taq DNA ligase and the initial DNA substrate strands. The incubation on ice prevented/avoided nonspecific ligation reactions during dilution. Note that ramps of the temperature changes are not included in the indicated times and are typically 3.3 K/s for both the heating and cooling when using aBio-Rad CFX96 Real-Time PCR thermal cycler.

| Step |   | Temperature (°C) | Period (s) |
|---|---|---|---|
| 1 | Initial denaturation | 95 | 20 |
| 2 | Denaturation | 95 | 5 |
| 3 | Annealing and ligation | 67 | 10 |
| 4 | GOTO step 2 (40 or 50 cycles) |   |   |
| 5 | Heating | 80 | 1 |
| 6 | Rapid cooling | 0 | - |
| 7 | Serial dilution |   |   |
| 8 | GOTO step 1 |   |   |

## S2. Quantitative PCR (COLD PCR)

A modified real-time PCR protocol referred in the main text as "COLD PCR" was implemented to assess the sequence (order) arrangement of the oligonucleotide DNA strands resulting from the ligation reaction experiments (Figure 2a-c, 4b, 5a-c, and 6a[S3]). The COLD PCR method was intended to detect and amplify 40-base sequences (AB, AC, BA, BC, CA, CB) under low denaturing temperature (Figure S3a and Table S3). PCR cycles were performed on a thermal cycler (Bio-Rad, CFX96 Touch™ Real-Time PCR, USA) with a reaction volume of 2.4 µl containing 0.02 units/µl Hotstart Phusion Polymerase (New England Biolabs, MA, USA), 1x detergent-free HF Phusion buffer (New England Biolabs), 200 µM total dNTP mixture (New England Biolabs), 1x EvaGreen (Biotium, CA, USA), and 125 nM of each forward and reverse primers. The Phusion polymerase was used since the polymerase lacks 5' -> 3' exonuclease activity. The PCR reaction solution was prepared on 96-well plates by an ultrasonic-based pipetting robot (LabCyto, USA).

40-base sequences were amplified with the following forward-primers sequences $a_{PCR}$: GCCAT AAAGG TGGAA GTGCT GGTT, $b_{PCR}$: GCCAT AAAGG GACAA GGCAA CAGT, $c_{PCR}$: GCCAT AAGGG TCACA TCGGA GTCT and reverse-primer sequences $\bar{a}_{PCR}$: GCGTA TTCCA GCACT TCCAC CTGA, $\bar{b}_{PCR}$: GCGTA TTTGT TGCCT TGTCC CTCA, and $\bar{c}_{PCR}$: GCGTA TTACT CCGAT GTGAC CCAA. Note that three bases at the 5' end were removed and short sequences GCCATAA (for forward primers) or GCGTATT (for reverse primers) were instead included in the primer sequences. These 5' short modified sequences were not complementary to the ligation products and increased the specificity of PCR amplification. Primer pairs $a_{PCR}$ and $\bar{b}_{PCR}$ were used to amplify and detect the AB sequence, whille $b_{PCR}$ and $\bar{c}_{PCR}$ were used for the BC sequence.

Note that in Figure 2a, b, c and S5, in order to differentiate *ab* and *abc* by PCR, we used instead of *b* a longer DNA substrate sequence, *bq* = ATGAG GGACA AGGCA ACAGT GAACT CAGTG TAGCC TCAGAT for the ligation reactions. We used higher denaturing temperature (98 °C) for all the cycles with the primers $a_{PCR}$ and $q_{PCR}$ = GCGTA TTTGA GGCTA CACTG AGTTC to detect BC or $a_{PCR}$ and $\bar{c}_{PCR}$, to detect ABC. We measured the amounts of the 40-base motifs by determining the $C_t$ values with the maxRatio method[S4]. See Figure S3b for the PCR curves and S3c and S3d for the estimated $C_t$ values for calibration, where the real time PCR cycles were initiated by controlled template concentrations. We fitted these $C_t$-value series by logarithmic curves $[\text{Template}] = 2^{-\alpha(C_t - \beta)}$ and obtained the calibration parameters $\alpha$ and $\beta$ (Table S4).

COLD PCR

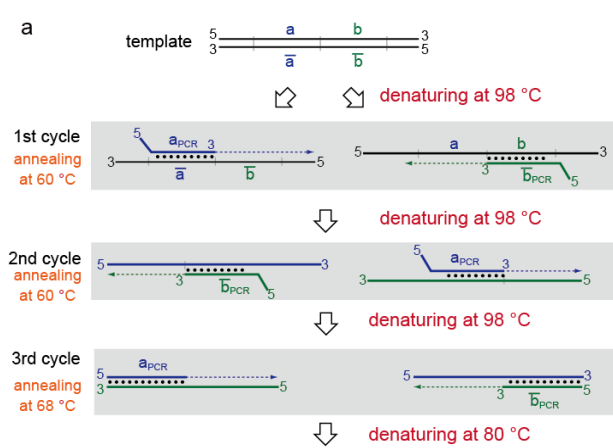

Calibration

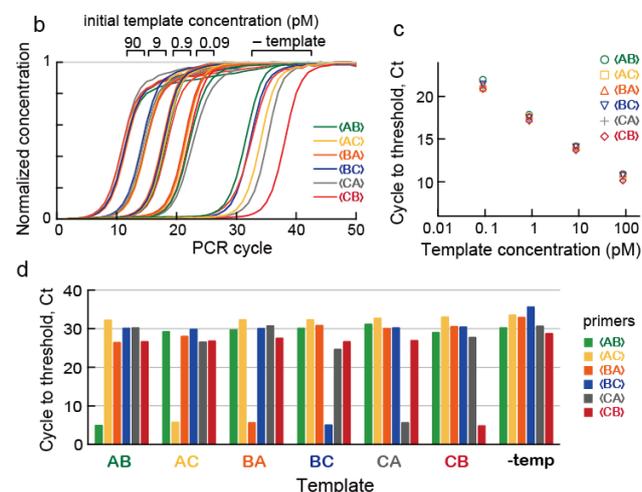

*Figure S3: COLD PCR method. **a**, By denaturing at a low temperature, only short strands of dimeric sequence motifs were amplified, confirmed by a melting curve analysis after the amplification. In the first three cycles, the denaturing temperature was high to denature ligation products, which possibly contain long double strands. **b**, Calibration curves of COLD PCR. We initiated the COLD PCR cycles by dimeric strands mixture containing AB, AC, BA, BC, CA, and CB or without templates of different concentrations. PCR cycles were performed with different primers. **c**, We analyzed the horizontal positions of the PCR curves, or the cycle to threshold denoted by $C_t$, by locating the cycling number where the ratio of the concentrations between two successive cycles becomes maximum[S2]. Let $x_n$ be the strand concentration at the n-th cycle. $C_t$ is n that maximizes the ratio $x_n / x_{n-1}$. By fitting logarithmic curves to the $C_t$-value series, we obtained calibration parameters (Table S4). **d**, Nonspecific amplification was checked by measuring the $C_t$ values of the COLD PCR cycles initiated by different dimeric template (1 nM) and amplified different primer sets. See Table S3 for the protocol of COLD PCR.*

*Table S3: COLD PCR protocol.* Only in the nonspecificity test (Figure S3d), the annealing temperature in step 7 was changed to 67 °C, but this small difference had only a small effect on the cycles to threshold values of PCR amplifications.

| Step | | Temperature (°C) | Period (s) |
|---|---|---|---|
| 1 | Initial denaturation | 98 | 60 |
| 2 | Denaturation | 98 | 5 |
| 3 | Annealing | 60 | 15 |
| 4 | Elongation and measurement | 72 | 5 |
| 5 | GOTO 2 (2 cycles) | | |
| 6 | Cold denaturation | 80 | 15 |
| 7 | Annealing | 68 | 15 |
| 8 | Elongation and measurement | 72 | 5 |
| 9 | GOTO 6 (50 cycles) | | |

*Table S4*: Calibration parameters of COLD PCR, $C_t$ values, obtained by fitting curves $[\text{Template}] = 2^{-\alpha(C_t - \beta)}$ *(in the unit of nM) to the $C_t$ values in Figure S3b.*

| Sequence motif | AB | AC | BA | BC | CA | CB |
|---|---|---|---|---|---|---|
| α | 0.90 | 0.98 | 0.96 | 0.94 | 0.94 | 0.93 |
| β | 17.8 | 17.3 | 17.6 | 17.5 | 17.0 | 17.0 |

## S3. Dissociation constant and ligation rate

In order to measure the dissociation constants $K_D$ and the ligation rate $k$ (Figure S4), we performed the ligation reactions under thermal cycling and without serial dilution (Figure S5). A solution initially containing the oligonucleotide strands $a$, $b$ and the complementary template $\overline{cba}$ (Figure S5a) could give rise to the product strand $ab$ upon hybridization of the substrate oligonucleotides $a$ and $b$ on the template $\overline{cba}$ and a subsequent ligation reaction. However, the newly formed $ab$ strand cannot template other ligation reactions in the absence of the complementary sequences $a$ and $b$. The amount of template strands in this case would does increase linearly. By comparing the ligations $a + b \rightarrow ab$ and $a + bc \rightarrow abc$ on the template $\overline{cba}$, we obtain $K_D$ and $k$ as follows.

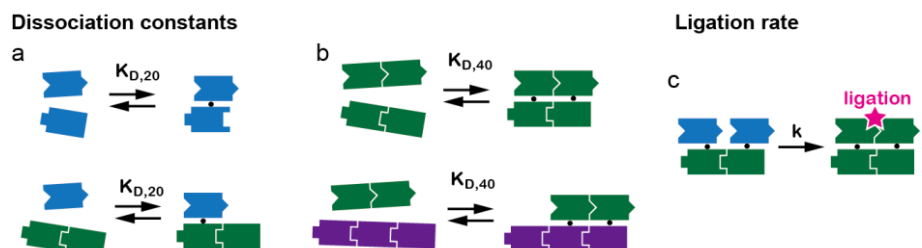

**Figure S4**: Dissociation constant $K_D$ and ligation rate $k$. **a**, **b**, $K_{D,20}$ and $K_{D,40}$ correspond to the dissociation constants of hybridization of 20-base or 40-base DNA oligonucleotide strands, respectively. $K_D$ values are inversely proportional to the stability of hybridization between complementary strands **c**, The ligation rate k was defined as the ligation reaction rate of two DNA oligonucleotide strands hybridized on the same template strand.

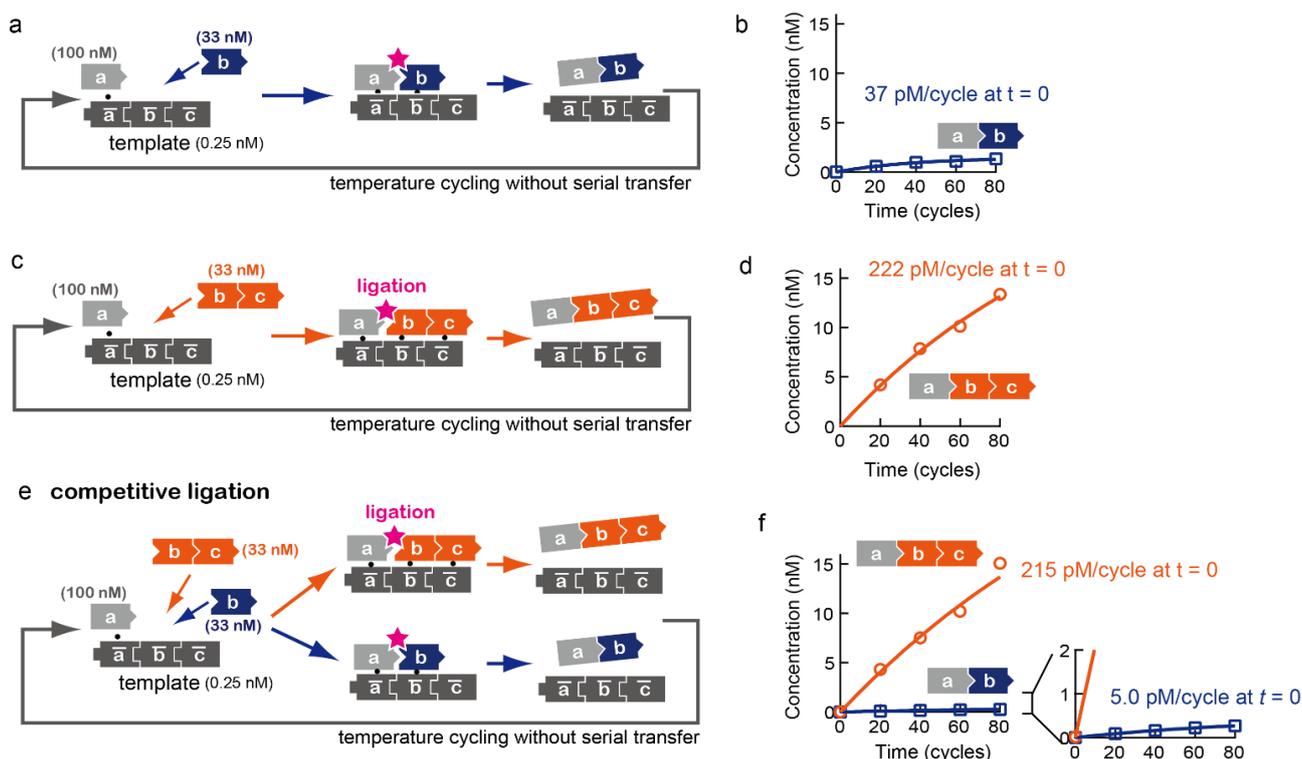

**Figure S5**: Linear ligation experiments for measuring the dissociation constants $K_D$ the reaction rate $k$ with various combinations of substrates. We performed thermal cycles between 67 °C for 10 s and 95 °C for 5 s without serial dilutions. To measure the amount of AB and ABC /independentlyseparately, we used a longer substrate bq = ATGAG GGACA AGGCA ACAGT GAACT CAGTG TAGCC TCAGA T for b. The sequence q does not hybridize to c and is floating when b hybridizes to b. See supplement S2 for the probing. The amount of each sequence was measured by COLD PCR[S3]. The solid curves were obtained by simulation. **a**, **b**, Substrates are a and b. The ligation rate was 37 pM/cycle. **c**, **d**, Substrates are a and bc. The ligation rate was 222 pM/cycle. **e**, **f**, Substrates are a, b, and bc (competitive binding). The ligation rates of ABC and AB were 215 and 5.0 pM/cycle, respectively. We found that (i) dimers are ligated much faster than monomers. (ii) When dimers and monomers compete, ligation of monomers are suppressed substantially while the ligation of dimers are not significantly affected. 40-base strands are ligated about 43 times faster than monomers. Solid curves in **b**, **d**, and **f** are curves fitted by $y(t) = (v_0 / q) [1 - \exp(-qt)]$ with fitting parameters $v_0$ and $q$.

## S4. Gel electrophoresis

Denaturing electrophoresis[S5] was run at 400 V for 15 min in 1x Tris-Borate-EDTA buffer (pH 8) with denaturing polyacrylamide gels containing 12.5% acrylamide/bisacrylamide (19:1) and 50% urea (Carl Roth, Karlsruhe, Germany or WAKO, Japan). DNA strands were detected with a fluorescent dye 1x SYBR Gold (Thermo Fisher Scientific, MA, USA). The gel image was taken by a CCD camera on an UV transilluminator (Figure S6) or an LED-illuminated gel documentation station (Figure S7a) and analysed by lab-developed LabVIEW (National Instruments, USA) program. The pixel intensity data was projected to one dimensional intensity profile in the lane axis direction. Peaks in the profile were fitted by super-positions of Lorentzian functions to estimate the DNA amount contained in each band (Figure 4a, 6b, and S7b-e).

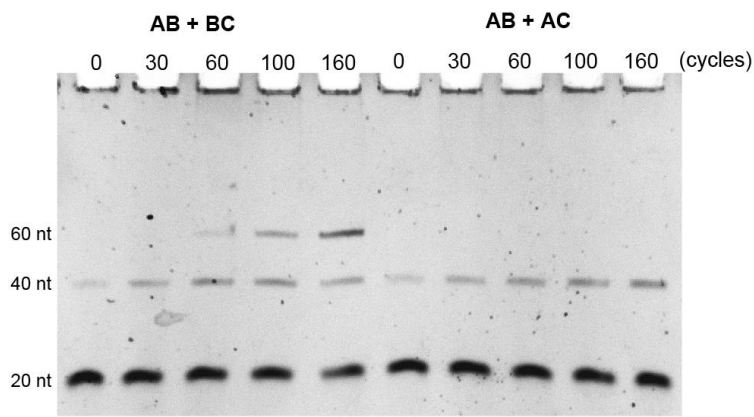

*Figure S6: Measurement of the DNA-strand length distribution. **a**, **b**, Electrophoresis gels stained with 1x SYBR Gold. The data correspond to the length distributions in Figure 6b.*

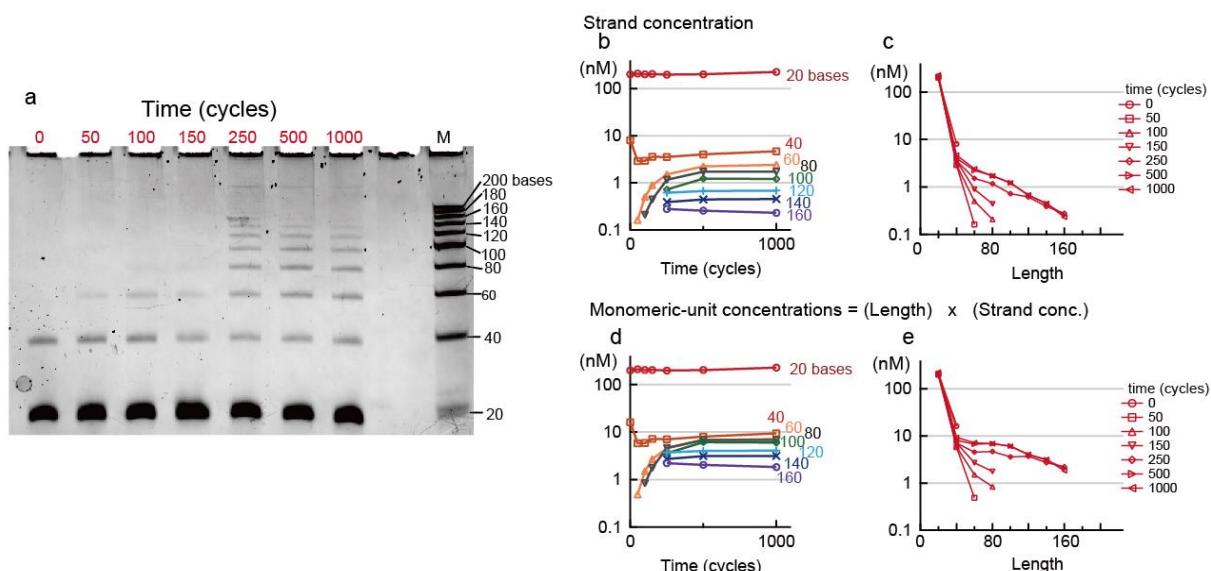

*Figure S7: Measurement of length distribution of the DNA strands. **a**, Electrophoresis gels stained with 1x SYBR Gold. The data correspond to the length distributions in Figure 6b. **b**, **c**, Strand concentrations. **d**, **e**, Monomeric-unit concentrations indicate how much monomer molecules are found in a given oligomer length.*

## S5. Numerical simulations

### 5.1 Cooperative ligation (Figure 4).

The cooperative ligation experiments in Figure 4 were modeled with relatively simple equations. The maximun length that could be obtained in our ligation reaction experiments was of 60nt DNA strands due to the limited number of 40nt DNA strands supplemented in the reactions.Because the strands do not elongate longer than 60 bases due to limited number of initial 40-base template strands, the model equations are relatively simple. Experimentally this is ensured by the lack of terminal phosphates for ligation. To test and for completeness, not only the three-body binding, but also four-body bindings were calculated. Including these computationally very elaborate, but in the reaction not very significant complexes did not affect the results significantly. Apparently, they are well enough covered by the modeling of the dimeric binding kinetics. The solved equations in the Mathematica (Wolfram) form is shown below.

```
(* Reactions *)

a0'[t] == k[t] (-a[t]b[t]ab[t] -a[t]bc[t]ab[t] -a[t]c[t]ac[t] -a[t]b[t]abc[t] -a[t]bc[t]abc[t]*r2)*r1*r1 +(-
a[t]b[t]c[t]abc[t] - ab[t]c[t]a[t]bc[t])*r1*r1*r1,

b0'[t] == k[t] (-a[t]b[t]ab[t] -b[t]c[t]bc[t] -a[t]b[t]abc[t] -b[t]c[t]abc[t])*r1*r1+(-
2a[t]b[t]c[t]abc[t] )*r1*r1*r1 - db0[t],

c0'[t] == k[t] (-a[t]c[t]ac[t] -b[t]c[t]bc[t] -ab[t]c[t]bc[t] -b[t]c[t]abc[t] -ab[t]c[t]abc[t]*r2)*r1*r1 +(-
a[t]b[t]c[t]abc[t] - a[t]bc[t]ab[t]c[t])*r1*r1*r1,

ab0'[t] == k[t] (+a[t]b[t]ab[t] -ab[t]c[t]bc[t] +a[t]b[t]abc[t] -ab[t]c[t]abc[t]*r2)*r1*r1 + (a[t]b[t]c[t]abc[t] -
a[t]bc[t]ab[t]c[t])*r1*r1*r1,

ac0'[t] == k[t] (+a[t]c[t]ac[t])*r1*r1,

bc0'[t] == k[t] (-a[t]bc[t]ab[t] +b[t]c[t]bc[t] -a[t]bc[t]abc[t]*r2 +b[t]c[t]abc[t])*r1*r1 + (a[t]b[t]c[t]abc[t] -
ab[t]c[t]a[t]bc[t])*r1*r1*r1,

abc0'[t] == k[t] (+a[t]bc[t]ab[t] +ab[t]c[t]bc[t] +a[t]bc[t]abc[t]*r2 +ab[t]c[t]abc[t]*r2)*r1*r1 +
(a[t]bc[t]ab[t]c[t]+ab[t]c[t]a[t]bc[t])*r1*r1*r1,

(* Conservation laws *)

a0[t] == a[t] +(+a[t]a[t] +a[t]ab[t] +a[t]ac[t] +a[t]abc[t])*r1 +
(+a[t]b[t]ab[t]+a[t]bc[t]ab[t]+a[t]c[t]ac[t]+a[t]b[t]abc[t]+a[t]c[t]abc[t]+a[t]ac[t]abc[t]+a[t]bc[t]abc[t]*r2)*r1*
r1 +(a[t]b[t]c[t]abc[t] +ab[t]c[t]a[t]bc[t])*r1*r1*r1,

b0[t] == b[t] +(+b[t]b[t] +b[t]ab[t] +b[t]bc[t] +b[t]abc[t])*r1 +
(+a[t]b[t]ab[t]+b[t]c[t]bc[t]+a[t]b[t]abc[t]+b[t]c[t]abc[t])*r1*r1,

c0[t] == c[t] +(+c[t]c[t] +c[t]ac[t] +c[t]bc[t] +c[t]abc[t])*r1 +
(+a[t]c[t]ac[t]+b[t]c[t]bc[t]+ab[t]c[t]bc[t]+a[t]c[t]abc[t]+b[t]c[t]abc[t]+ab[t]c[t]abc[t]*r2+ac[t]c[t]abc[t])*r1*
r1 +(a[t]b[t]c[t]abc[t] +a[t]bc[t]ab[t]c[t])*r1*r1*r1,

ab0[t] == ab[t] +(+a[t]ab[t] +b[t]ab[t] +ab[t]ab[t]*r2 +ac[t]ab[t] +bc[t]ab[t] +ab[t]abc[t]*r2)*r1 +
(+ab[t]c[t]bc[t]+a[t]b[t]ab[t]+a[t]bc[t]ab[t]+ab[t]c[t]abc[t]*r2)*r1*r1+a[t]bc[t]ab[t]c[t]*r1*r1*r1,

ac0[t] == ac[t] +(+a[t]ac[t] +c[t]ac[t] +ab[t]ac[t] +ac[t]ac[t]*r2 +bc[t]ac[t] +ac[t]abc[t] +ac[t]abc[t])*r1 +
(+a[t]c[t]ac[t]+a[t]ac[t]abc[t]+ac[t]c[t]abc[t])*r1*r1,

bc0[t] == bc[t] +(+b[t]bc[t] +c[t]bc[t] +ab[t]bc[t] +ac[t]bc[t] +bc[t]bc[t]*r2 +bc[t]abc[t]*r2)*r1 +
(+a[t]bc[t]ab[t]+b[t]c[t]bc[t]+ab[t]c[t]bc[t]+a[t]bc[t]abc[t]*r2)*r1*r1+ab[t]c[t]a[t]bc[t]*r1*r1*r1,

abc0[t] == abc[t] +(+a[t]abc[t] +b[t]abc[t] +c[t]abc[t] +ab[t]abc[t]*r2 +ac[t]abc[t] +ac[t]abc[t] +bc[t]abc[t]*r2
+abc[t]abc[t]*r3)*r1 +
(+a[t]b[t]abc[t]+a[t]c[t]abc[t]+a[t]ac[t]abc[t]+a[t]bc[t]abc[t]*r2+b[t]c[t]abc[t]+ab[t]c[t]abc[t]*r2+ac[t]c[t]abc[
t])*r1*r1+ a[t]b[t]c[t]abc[t])*r1*r1*r1
```

In this case, a[t], b[t]… corresponded to the concentrations of the free oligonucleotide strands while a0[t], b0[t]… indicated the total concentration (including hybridized) of the oligonucleotide strands. r1 = 1/$k_{D,20}$., r2 = $k_{D,20}$/$k_{D,40}$, r3 = $k_{D,20}$/$k_{D,60}$. The ligation rate k[t] wask0 exp[- t / 80] with k0 = 3.0 nM/s.

For the simulation curves in Figure 4b, we plotted <AB> = 2(ab0[t] + abc0[t]) , <BC> = 2(bc0[t] + abc0[t]), and <AC> = 2ac0[t] for the simulation curves in "PCR" plots, and monomers = 2(a0[t] + b0[t] + c0[t]), dimers = 2(ab0[t] + bc0[t] + ac0[t]), and trimers = 2abc0[t] for the simulation curves in "Gel" plots.

**5.2 Selection by cooperation (Figure 5).**

As explained in the previous section, the strands do not elongate longer than 60 bases due to limited number of initial 40-base template strands, the model equations are relatively simple. Again, in experiment this is ensured by the lack of terminal phosphates which inhibit the elongation during ligation. Again, to test for completeness, not only the three-body binding, but also four-body bindings were calculated. Including these computationally very elaborate, but in the reaction not very significant complexes did not affect the results significantly. Apparently, they were already enough covered by the modeling of the dimeric binding kinetics.

In Figure 5d, the $k_D$ parameter was biased/changed in a way that A and B could hybridize more stably than C. More specifically, $k_D$ was biased as $\sqrt{m}k_D$ for the hybridization including A and B, and $\sqrt{1/m}\,k_D$ for the hybridization including C. Here, the parameter $m$ is the number indicated in Figure 5d as 1.3, 2, 3, and 4 fold. If the hybridization contains multiple units, we multiply all the biases. For example, the $k_D$ of a complex with the hybridization of A and BC on ABC, the dissociation constant of this complex becomes $\sqrt{m}\sqrt{m}\sqrt{1/m}\,k_{D,20}k_{D,40} = \sqrt{m}k_{D,20}k_{D,40}$.

**5.3 Domination by the majority sequence motifs (Figure 6a).**

We used the additional symmetries such as [AB] = [BC] = [CA] and [CB] = [BA] = [AC] to reduce the computational time without affecting the results. Since the number of possible sequences grew/increased exponentially with the length, we restricted the maximum length to six and neglected the reactions that produced DNA strands longer than 120nt nucleotides. As shown before, the four strand complexes which were computationally very expensive were not expected to significantly affect the results and were therefore not calculated.

**5.4 Length distribution (Figure 6b).**

The simulation results indicated that the length distribution was mostly determined by the dominant sequences. Thus, considering only reactions including sequence motifs AB, BC, and CA, the number of reaction terms decreased dramatically and the prediction of the length distribution was not significantly affected. The length distribution was calculated by using terms of a length up to 640 bases (32 units).

**5.5 Symmetry breaking (Fig. 7a) and spatial pattern formation (Fig. 7b,c)**

Simulations in Figure S8 and Figure 7 were conducted with initial uniform concentrations of all nine dimer templates with 5% concentration fluctuations. The maximum length was restricted to 80 bases and the reactions producing strands longer than 80 bases were neglected in order to reduce the computational time. Moreover, four strand binding complexes were not considered. A slightly higher value of he dilution rate($d$ = 0.045 /cycle) compared to the one used for the experiments was used to better match the experimentally found splitting of the concentration of the two competing sequences.

## S6. Stochastic emergence of self-sustaining sequence structure

For the one-dimensional simulations in Figure 7b,c, diffusion terms such as $D\partial^2[A]/\partial x^2$ were incorporated into all the reaction equations. For simplicity, *D* indicated a diffusion coefficient and it was not depending on the sequence nor the strand length. We used a periodic boundary condition. In the simulations, we find the following six patterns of collaboration with the used three sequences if the cooperation is active:

- Periodic patterns with three letters:

    AB,BC,CA → ...ABC...
    AC,BA,CB → ...BCA...

- Periodic patterns with two letters and homogeneous sequences:

    BC,CB,AA → ...BCBC... and ...AAA...
    AC,CA,BB → ...ACAC... and ...BBB...
    AB,BA,CC → ...ABAB... and ...CCC...

- The completely homogeneous case:

    AA,BB,CC → ...AAA..., ...BBB... and ...CCC...

Below in Figure S8, we present simulation runs with a homogeneously mixed solution. As can be seen, the system amplifies the stochastic initial conditions with 5% variations in concentration to emerge above cooperating two letter sequences. We find the six classes of cooperation as indicated above.

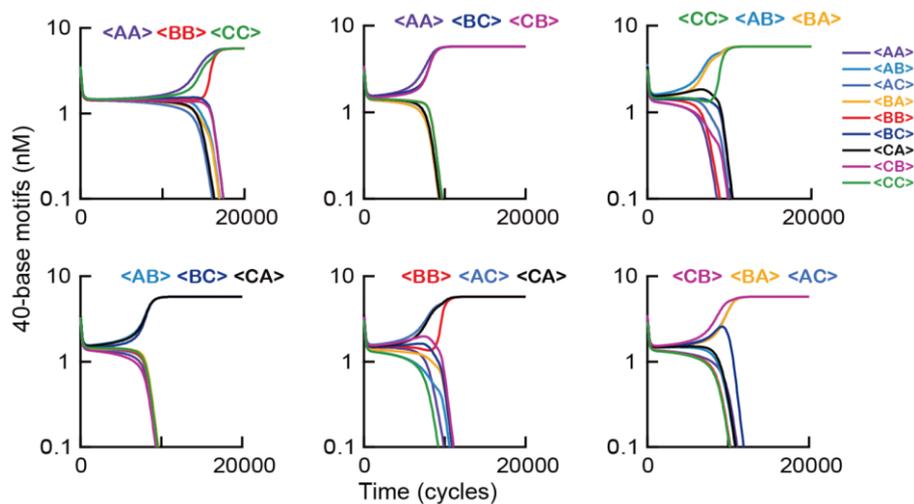

*Figure S8:* Stochastic emergence of sequence structures was observed when reactions were initiating with a uniform concentration of all the nine possible dimers (AA, AB, …, CC) considering 5% concentration fluctuations. We observed a state dominated by one of six possible cooperative combinations. For example, AB and AC did not coexist because AB and AC were competing; B and C compete to get the right hand side of A. The dilution rate was d = 0.045 / cycle.

## S7. Theoretical analysis

The mechanism of the higher-order growth by the cooperation and the frequency-dependent separation was theoretically studied here.

### 7.1 Competition between two sequence motifs

In the Appendix section, a simple rate equation (14) for the competition between two sequence motifs $\alpha$ and $\beta$ was derived. The equation (14) is copied from the main manuscript for convenience by substituting $k'$ by $k$ for simplicity.

$$\begin{pmatrix}\dot{\alpha}\\ \dot{\beta}\end{pmatrix} = k(S-\alpha-\beta)^2 \begin{pmatrix}(1+pL_\alpha)\alpha\\ (1+pL_\beta)\beta\end{pmatrix} - d\begin{pmatrix}\alpha\\ \beta\end{pmatrix} \qquad (7.1.1)$$

As a comprehensive example, let $\alpha$ and $\beta$ correspond to {<AB>, <BC>, and <CA>} and {<CB>, <BA>, and <AC>}, respectively, in Figure 6a. $S - \alpha - \beta$ indicated the amount of the substrates.

We did a standard linear-stability analysis to study the equations. The steady state was obtained by solving $\alpha_0[k(S-\alpha_0-\beta_0)(1+pL_{\alpha_0})-d] = 0$ and $\beta_0[k(S-\alpha_0-\beta_0)(1+pL_{\beta_0})-d] = 0$. There were three possible solutions;

(i) $\alpha_0 = \beta_0 = 0$ (Trivial fixed point),

(ii) $\{\alpha_0 = 0, \beta_0 \neq 0\}$ or $\{\alpha \neq 0, \beta = 0\}$ (Splitting), and

(iii) $\alpha_0 = \beta_0 \neq 0$ (Uniform state).

The stability of these steady states against a small perturbation were determined by the sign of the eigenvalues of the Jacob matrix. If all the eigenvalues were negative, the state was stable. Inversely, for positive values of the eigenvalues, the state was found to be unstable. The Jacob matrix at the steady state wsa

$$J(\alpha_0,\beta_0) = \begin{pmatrix}\partial_\alpha f_\alpha(\alpha_0,\beta_0) & \partial_\beta f_\alpha(\alpha_0,\beta_0)\\ \partial_\alpha f_\beta(\alpha_0,\beta_0) & \partial_\beta f_\beta(\alpha_0,\beta_0)\end{pmatrix}$$

$$= k(S-\alpha_0-\beta_0)\begin{pmatrix}-2\alpha_0(1+pL_{\alpha_0}) + (S-\alpha_0-\beta_0)p\frac{\partial L_{\alpha_0}}{\partial \alpha_0} & -2\alpha_0(1+pL_{\alpha_0})\\ -2\beta_0(1+pL_{\beta_0}) & -2\beta_0(1+p\beta_0) + (S-\alpha_0-\beta_0)p\frac{\partial L_{\beta_0}}{\partial \beta_0}\end{pmatrix}$$

The eigenvalues of $J(\alpha_0,\beta_0)$ at the uniform state $\alpha_0 = \beta_0$ were $\lambda_1 = kp\alpha_0(S-2\alpha_0)^2\frac{\partial L_{\alpha_0}}{\partial \alpha_0}$ and $\lambda_2 = \lambda_1 - 4k(S-2\alpha_0)(1+pL_{\alpha_0})\alpha_0$. Because $\frac{\partial L_{\alpha_0}}{\partial \alpha_0} > 0$, the sign of the cooperativity $p$ determined the stability of the uniform state. If $p > 0$ (positive cooperation, hyperexponential growth), the uniform state was unstable because $\lambda_1 > 0$. Separation takes place. If $p = 0$ (no cooperation, exponential growth), uniform state is neutral because $\lambda_1 = 0$ and $\lambda_2 = -4k(S-2\alpha_0)\alpha_0 < 0$. If $p < 0$ (negative cooperation, subexponential growth), uniform state can be stable. Such negative cooperation was possible if the product inhibition was strong. The motifs with lower frequency had higher growth rates. The simulated curves by (7.1.1) are shown in Figure 8d. $L_\alpha$ was approximated by $\alpha$ for simplicity.

### 7.2 One dimensional system

To discuss the pattern formation observed in the simulation (Figure 7b,c), we added a diffusional term to (7.1.1).

$$\begin{pmatrix}\dot{\alpha}\\ \dot{\beta}\end{pmatrix} = k(S-\alpha-\beta)^2\begin{pmatrix}(1+pL_\alpha)\alpha\\ (1+pL_\beta)\beta\end{pmatrix} - d\begin{pmatrix}\alpha\\ \beta\end{pmatrix} + D\frac{\partial^2}{\partial x^2}\begin{pmatrix}\alpha\\ \beta\end{pmatrix} \qquad (7.2.1)$$

Here, $D$ indicated a diffusion coefficient. The linearized equation system is

$$\frac{\partial}{\partial t}\begin{pmatrix}\alpha\\ \beta\end{pmatrix} = J(a_0,b_0)\begin{pmatrix}\alpha\\ \beta\end{pmatrix} + D\frac{\partial^2}{\partial x^2}\begin{pmatrix}\alpha\\ \beta\end{pmatrix}.$$

By inserting the perturbation around the steady state

$$\begin{pmatrix}\Delta_\alpha(x,t)\\ \Delta_\beta(x,t)\end{pmatrix} \equiv \begin{pmatrix}\alpha(x,t)-\alpha_0\\ \beta(x,t)-\beta_0\end{pmatrix} = e^{\lambda_q t + iqx}\begin{pmatrix}\alpha_0\\ \beta_0\end{pmatrix},$$

the following equation was obtained

$$\lambda_q\begin{pmatrix}\alpha_0\\ \beta_0\end{pmatrix} = \left[J(\alpha_0,\beta_0) - q^2 D\begin{pmatrix}1 & 0\\ 0 & 1\end{pmatrix}\right]\begin{pmatrix}\alpha_0\\ \beta_0\end{pmatrix}.$$

The eigenvalues of $J(\alpha_0,\beta_0) - q^2D\begin{pmatrix}1 & 0\\ 0 & 1\end{pmatrix}$ were $\lambda_1 - Dq^2$ and $\lambda_2 - Dq^2$. The sign of these eigenvalues determined the stability of the uniform state and varied depending on the wave number $q$. When $p > 0$, the stability of the uniform state was determined by the sign of $\lambda_1 - Dq^2 = kp\alpha_0(S - 2\alpha_0)^2 \frac{\partial L_{\alpha_0}}{\partial \alpha_0} - Dq^2$ because $\lambda_1$ was always larger than $\lambda_2$ as shown in section 7.1. In the wave-number range of $q < q_{\max} \equiv (S - 2\alpha_0)\sqrt{\frac{kp\alpha_0}{D}\frac{\partial L_{\alpha_0}}{\partial \alpha_0}}$, a uniform state was unstable and a spatial pattern formed spontaneously. This inequality reasonably suggested that the pattern became finer with a smaller diffusion and a larger reaction rate. Figure S9 showed an example of the pattern formation simulated by (7.2.1) with different $p$. $L_\alpha$ was approximated by $\alpha$ for simplicity. This approximation demonstrated that the pattern formation did not take place without cooperation (Figure S9b) as shown in Figure 7c.

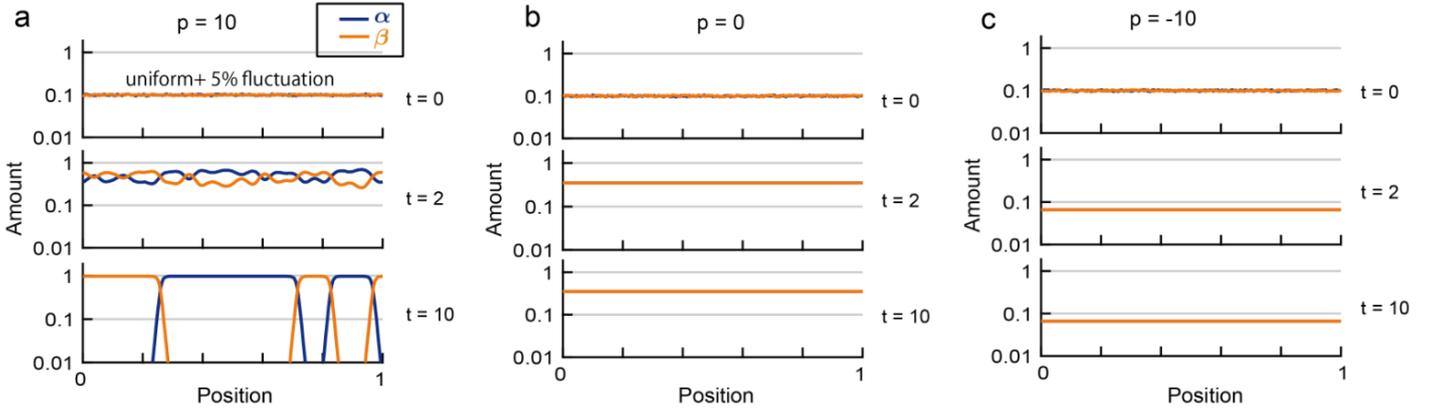

**Figure S9**. The equations (7.2.1) with an approximation of $L_\alpha = \alpha$ for simplicity were numerically solved with the parameters k = 1, S = 1, d = 0.5, and different values of p indicated. The initial conditions were 0.1 with 5% fluctuations picked from a uniform distribution. The snapshots of the concentration profiles of $\alpha(x,t)$ and $\beta(x,t)$ are shown at t = 0, 2, 10. **a.** If p > 0, the uniform state resulted to be unstable, and a spatial pattern formed spontaneously. **b.** If p = 0, the stability was neutral as in (7.1). However, the diffusion flattened the concentration profile. **c.** If p < 0, the uniform state was stable.

## S8. Supplementary experimental data

Replicates of experimental data of Figure 4, 5b, c and 6a. are shown in Figure S10 and demonstrate the robustness of the effect.

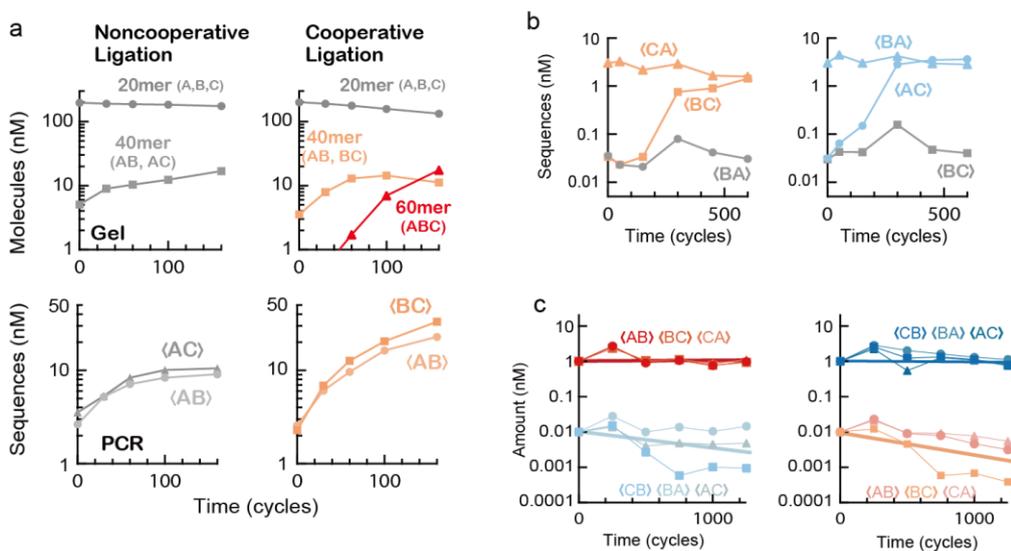

*Figure S10*. Replicate experiments of Figure 4, 5b,c, and 6a. (*a*) shown is a replicate of Figure 4, conducted with an initial concentration of the templates of 2 nM. (*b*) Replicate of Fig. 5b and c, but the initial concentrations are 2 nM for CA (left) and BA (right) and 0.02 nM for BC and BA (left) and AC and BC (right). (*c*) Replicate of Figure 6a, but the initial concentrations of the templates are 1 nM for the major sequences and 0.01 nM for minor sequences. The dilution rate is 0.47 / s.

## References for supplement